\documentclass[11pt]{article}

\usepackage[preprint]{acl}

\usepackage{adjustbox}
\usepackage{amsmath}

\usepackage{amssymb} %
\usepackage{array}
\usepackage{bbding}
\usepackage{booktabs}

\usepackage{colortbl}
\usepackage{enumitem}
\usepackage{float}
\usepackage{graphicx}
\usepackage{anyfontsize}
\usepackage{longtable}
\usepackage{makecell}
\usepackage{mdframed}
\usepackage{multirow}
\usepackage{natbib}
\usepackage{placeins}
\usepackage{subcaption}
\usepackage{siunitx} 
\usepackage{xcolor}

\usepackage{pgfplots}
\usepackage{times}
\usepackage{latexsym}
\usepackage{colortbl}
\usepackage[T1]{fontenc}

\usepackage[utf8]{inputenc}

\usepackage{microtype}

\usepackage{inconsolata}

\usepackage{booktabs}
\usepackage{caption}

\usepackage{hyperref}
\usepackage{url}
\usepackage{wrapfig}   %
\usepackage{booktabs}  %
\usepackage{tikz}
\usepackage{pgfcalendar}
\usepackage{etoolbox}
\usepackage{xparse}
\usepackage{graphicx}    %
\usepackage{subcaption}
\usepackage{wasysym}
\usetikzlibrary{positioning, arrows.meta}
\usepackage{minitoc}
\usepackage{makecell}
\usepackage{xcolor, color, soul}

\usepackage{amsmath,amsfonts,bm}

\def\eqref#1{equation~\ref{#1}}

\def\1{\bm{1}}

\DeclareMathAlphabet{\mathsfit}{\encodingdefault}{\sfdefault}{m}{sl}
\SetMathAlphabet{\mathsfit}{bold}{\encodingdefault}{\sfdefault}{bx}{n}

\usepackage{enumitem}
\usepackage{pifont}
\usepackage{tikz}
\usepackage{xcolor}
\usepackage{xspace}
\usepackage{stfloats} %

\usepackage[most]{tcolorbox}
\usepackage{setspace} %
\definecolor{customblue}{HTML}{2E5AA7}

\usepackage{hyperref}
\hypersetup{
  colorlinks   = true,    %
  urlcolor     = customblue,    %
  linkcolor    = customblue,    %
  citecolor    = customblue,     %
  breaklinks   = true,
}

\usepackage[nameinlink, capitalise]{cleveref}       %

\newcommand{\eg}{\emph{e.g}.\xspace}
\newcommand{\ie}{\emph{i.e}.\xspace}

\usepackage{cancel}
\usepackage{bigdelim, rotating}  %

\newcommand{\mytoprule}[1]{\noalign{\vskip -\aboverulesep}\cmidrule[\heavyrulewidth]{#1}}
\newcommand{\mybottomrule}[1]{\cmidrule[\heavyrulewidth]{#1}\noalign{\vskip -\belowrulesep}}
\newcommand{\mymidrule}[1]{%
    \noalign{\vskip -\aboverulesep} 
    \cmidrule[\heavyrulewidth]{#1}  
    \noalign{\vskip -\belowrulesep} 
}

\newcommand{\revised}[1]{%
  \begingroup
  \renewcommand{\baselinestretch}{1}%
  \color{red}#1%
  \endgroup
}

\usepackage{etoolbox}
\robustify{\revised}

\definecolor{takeawaycolor}{RGB}{191, 1, 0}

\newcommand{\passk}{pass@$k$\xspace}
\newcommand{\passatk}[1]{pass@#1\xspace}

\newcommand{\originalstat}{$S_{orig}$\xspace}
\newcommand{\revisedstat}{$S_{rev}$\xspace}

\newcommand{\reasonforchange}{$R_r$\xspace}
\newcommand{\reasontext}{$R$\xspace}
\newcommand{\consequencetext}{$R_c$\xspace}

\newcommand{\revisionsummary}{$S_{rev}^{\prime}$\xspace}

\newcommand{\outlinerevtext}{\textit{outline-revision}\xspace}
\newcommand{\diffanalysistext}{\textit{reflect-revision}\xspace}
\newcommand{\fillcrtext}{\textit{discover-weakness}\xspace}
\newcommand{\outlinerevtitle}{Outline Revision\xspace}
\newcommand{\diffanalysistitle}{Reflect Revision\xspace}
\newcommand{\fillcrtitle}{Discover Weakness\xspace}

\newcommand{\outlinerevsymb}{$S_{orig} + R \to S_{rev}^{\prime}$\xspace}
\newcommand{\diffanalysissymb}{$\left| S_{rev} - S_{orig}\right| \to R$\xspace}

\newcommand{\fillcrsymb}{$S_{orig}\to R$\xspace}

\newcommand{\ourmodel}{\textbf{\textsc{CRitic}}-LLaMA-3.1-8B\xspace}
\newcommand{\critic}{\textbf{\textsc{CRitic}}\xspace}
\newcommand{\stageonedata}{\textbf{\textsc{CR-mix}}\xspace}
\newcommand{\stagetwodata}{\textbf{\textsc{CR-instruct}}\xspace}
\newcommand{\stagethreedata}{\textbf{\textsc{CR-sec}}\xspace}
\newcommand{\oureval}{\textbf{\textsc{CR-eval}}\xspace}

\usetikzlibrary{tikzmark}
\makeatletter
\newcommand*\myfontsize{%
\@setfontsize\myfontsize{6.7}{8}%
}
\makeatother

\definecolor{cadmiumgreen}{rgb}{0.0, 0.42, 0.24}

\definecolor{myred}{rgb}{0.7, 0.3, 0.0}
\definecolor{myblue}{rgb}{0.2, 0.3, 0.6}

\newenvironment{packeditemize}{
\begin{list}{$\bullet$}{
\setlength{\labelwidth}{6pt}
\setlength{\itemsep}{0pt}
\setlength{\leftmargin}{\labelwidth}
\addtolength{\leftmargin}{\labelsep}
\setlength{\parindent}{0pt}
\setlength{\listparindent}{\parindent}
\setlength{\parsep}{0pt}
\setlength{\topsep}{3pt}}}{\end{list}}

\newcolumntype{C}[1]{>{\centering\arraybackslash}p{#1}}

\NewEnviron{promptgroup}{%
  \begin{minipage}{\columnwidth}
    \BODY
  \end{minipage}
}

\newtcolorbox[auto counter]{promptbox}[2][]{
  float,
  float=htbp,  %
  width=\linewidth,
  colback=white,
  title={\fontsize{9.7}{7}\selectfont Prompt \thetcbcounter: #2},
  coltitle=black,
  left=5pt,
  right=5pt,
  top=5pt,
  bottom=5pt,
  fonttitle=\sffamily\bfseries\small,
  boxrule=1.2pt,
  label={#1},
  colframe=black,
  colbacktitle=white,
  before upper={\setstretch{0.9}},
  before={\par\vspace*{0pt}},
  after={\par\vspace*{0pt}},
}

\newtcolorbox[auto counter]{examplebox}[2][]{
  float,
  float=htbp,  %
  width=\linewidth,
  colback=white,
  title={\fontsize{9.7}{7}\selectfont Example \thetcbcounter: #2},
  coltitle=black,
  left=5pt,
  right=5pt,
  top=5pt,
  bottom=5pt,
  fonttitle=\sffamily\bfseries\small,
  boxrule=1.2pt,
  label={#1},
  colframe=black,
  colbacktitle=white,
  before upper={\setstretch{0.9}},
  before={\par\vspace*{0pt}},
  after={\par\vspace*{0pt}},
}

\crefname{figure}{Figure}{Figures}
\Crefname{figure}{Figure}{Figures}
\crefname{table}{Table}{Tables}
\Crefname{table}{Table}{Tables}
\crefname{equation}{Eq.}{Eqs.}
\Crefname{equation}{Eq.}{Eqs.}
\crefname{appendix}{Appendix}{Appendices}
\Crefname{appendix}{Appendix}{Appendices}

\makeatletter
\crefname{tcb@cnt@promptbox}{prompt}{prompts}    %
\Crefname{tcb@cnt@promptbox}{Prompt}{Prompts}    %
\makeatother

\makeatletter
\crefname{tcb@cnt@examplebox}{example}{examples}    %
\Crefname{tcb@cnt@examplebox}{Example}{Examples}    %
\makeatother

\usepackage{arydshln} %

\title{Can Large Language Models Automate the Refinement of\\ Cellular Network Specifications?}

\author{Jianshuo Dong, Yuanjie Li, Jun Liu, Hewu Li, Han Qiu$^{*}$ \\
Tsinghua University, China\\
\texttt{dongjs23@mails.tsinghua.edu.cn, qiuhan@tsinghua.edu.cn}
}

\begin{document}
\maketitle

\begin{abstract}
Cellular networks, \eg, 4G/5G, rely on complex technical specifications to ensure correct functionality; however, these specifications often contain flaws or ambiguities.
In this paper, we investigate the application of Large Language Models for \textit{automated cellular network specification refinement}.
We identify Change Requests, which record specification revisions, as a key source of domain-specific data and formulate specification refinement as three complementary sub-tasks.
We introduce \oureval, a benchmark of 200 security-related test cases, and evaluate 17 open-source and 14 proprietary models.
The best-performing model, \texttt{GPT-o3-mini}, identifies weaknesses in over 127 test cases within five trials.
We further study LLM specialization, showing that fine-tuning an 8B model can outperform advanced LLMs such as \texttt{DeepSeek-R1} and \texttt{Qwen3-235B}.
Evaluations on 30 real-world cellular attacks demonstrate the practical impact and remaining challenges.
The codebase and benchmark are available at \url{https://github.com/jianshuod/CR-Eval}.
\end{abstract}

\section{Introduction}

Recent advances in Large Language Models (LLMs)~\citep{achiam2023gpt-4-technical-report, GPT5-system-card, guo2025deepseek-r1} have sparked their remarkable applications across diverse domains, including finance~\citep{wu2023bloomberggpt-llm-for-finance-ref}, healthcare~\citep{singhal2023large-llm-for-medicine-ref-nature}, and mathematics~\citep{deepmind2024ai-solves-imo-problems-at-silver-medal-level}.
In this work, we investigate \textbf{\emph{automated cellular network specification refinement}}, a previously unthinkable yet now plausible concept with LLMs.

\begin{figure}[ht]
\centering
\includegraphics[width=\linewidth]{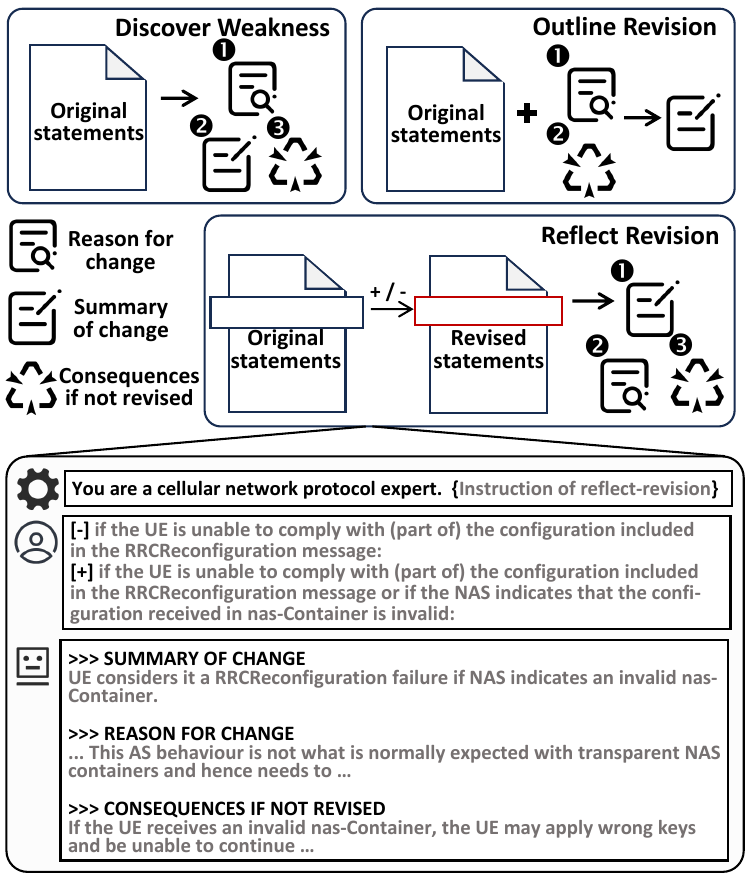}
\caption{\label{fig:task-formulation} Illustration of task formulation on \oureval.}
\end{figure}

Cellular networks, \eg, 4G/5G, rely on technical specifications to define protocol behavior and interoperability~\citep{gsma2025mobile-user-statistics, 3gpp-standard}. 
Despite their importance, these specifications often contain design flaws or ambiguities that can lead to security vulnerabilities or performance degradation~\citep{shaik2017practical-attacks-dos-ref-3-downgrading-ref-1}.
Detecting such weaknesses has traditionally relied on manual expert analysis~\citep{rupprecht2019breaking-lte-on-layer-2-dns-spoofing-and-website-spoofing-ref-1}, which is increasingly impractical as the standards rapidly expand\footnote{Measured by PDF page count, 3GPP standards grew from 59,258 pages in Release 8 (LTE) to 117,951 pages in Release 15 (5G), and to 195,752 pages in Release 18 as of March 2025.}. 
Existing automated approaches, including formal verification~\citep{hussain20195greasoner} and NLP-based methods~\citep{al2024hermes}, provide partial automation but still require expert effort and do not scale well with evolving specifications.
Consequently, the cellular networking community continues to call for more powerful and scalable automation to detect and mitigate specification weaknesses~\citep{3gpp.33.855-study-security-aspects-of-5g-networks, 3gpp2024security-assurance}. 
This motivates exploring LLMs for automated cellular specification refinement, given their strong language understanding and capable reasoning abilities.

To advance this, we identify and leverage 200,000+ approved Change Requests (CRs), which document historical specification revisions and necessary expert comments, as a data source.
Based on CRs, we devise three complementary LLM-tractable domain tasks: \fillcrtext (uncovering potential weaknesses in specifications), \outlinerevtext (proposing necessary revisions given weaknesses), and \diffanalysistext (ensuring revisions address weaknesses).
Under this framing, we propose \oureval, which comprises 200 security-related test cases and serves as a proxy for the real-world application of LLMs.
We extensively evaluate 17 open-source and 14 proprietary models, including \texttt{GPT-5}.
In the hardest yet imperative \fillcrtext task, the best-performing model \texttt{GPT-o3-mini} can discover weaknesses in over 127 out of 200 test cases within five trials.

We further explore domain specialization through fine-tuning.
We introduce an effective three-stage training recipe combined with a novel rationale augmentation technique.
The resulting domain-specialized 8B LLM nearly triples the performance of its base model (\texttt{LLaMA-3.1-8B}) on the \diffanalysistext task and even outperforms advanced models such as \texttt{Qwen3-235B} on the \fillcrtext task.
Further analysis of token-prediction behavior shows increased emphasis on security-related tokens and more precise technical terminology of cellular networks.
We conduct extensive experiments to examine the scalability of the training recipe and its extensibility to stronger base models.
In addition, we evaluate the domain-specialized LLM on 30 known cellular attacks, all of which are successfully detected.

As the incoming 6G technology integrates new features~\citep{lin20243gpp-5g-to-6g, gsma2025-6g-study-request}, it inevitably drives the evolution of 3GPP standards and raises concerns about new specifications.
We show that LLMs present a timely and effective opportunity for automated cellular specification refinement.
Our main contributions are threefold:
\begin{packeditemize}
    \item \textbf{New insight.} We pioneer LLM adoption for cellular specification refinement and strategically leverage change requests as domain data to form the foundation of a systematic study.
    \item \textbf{Evaluating LLMs' domain-specific ability.} We establish \oureval, which enables the community to understand the domain-specific abilities of modern LLMs. Using \oureval, we conduct an extensive measurement across 31 representative frontier LLMs.
    \item \textbf{Towards domain-specialized LLMs.} We explore avenues for domain specialization, including an effective fine-tuning recipe. We test on known cellular attacks to identify areas for further improvements in steering LLMs.
\end{packeditemize}

\section{Necessary Backgrounds}\label{sec:background}

\noindent \textbf{Largae Language Models (LLMs)}, for example, \texttt{GPT-5}~\citep{GPT5-system-card}, utilize the decoder-only Transformer architecture~\citep{vaswani2017attention-is-all-you-need} and are trained on the next-token-prediction task~\citep{radford2018gpt1} as:
{
\setlength{\abovedisplayskip}{3pt}
\setlength{\belowdisplayskip}{3pt}
\begin{equation}\label{eq:casual}
P(x_1,\cdots,x_N)=\prod_{i=1}^{N}P(x_i|x_1,\cdots,x_{i-1}),
\end{equation}}\noindent
where $P(x_i|x_1,\cdots,x_{i-1})$ represents the probability of predicting token $x_i$ given the preceding sequence $x_1,\cdots,x_{i-1}$, with tokens typically operating at the subword level.
During inference, LLMs can respond to user queries provided as prompts.

\begin{figure}[t]
    \centering
    \includegraphics[width=\linewidth]{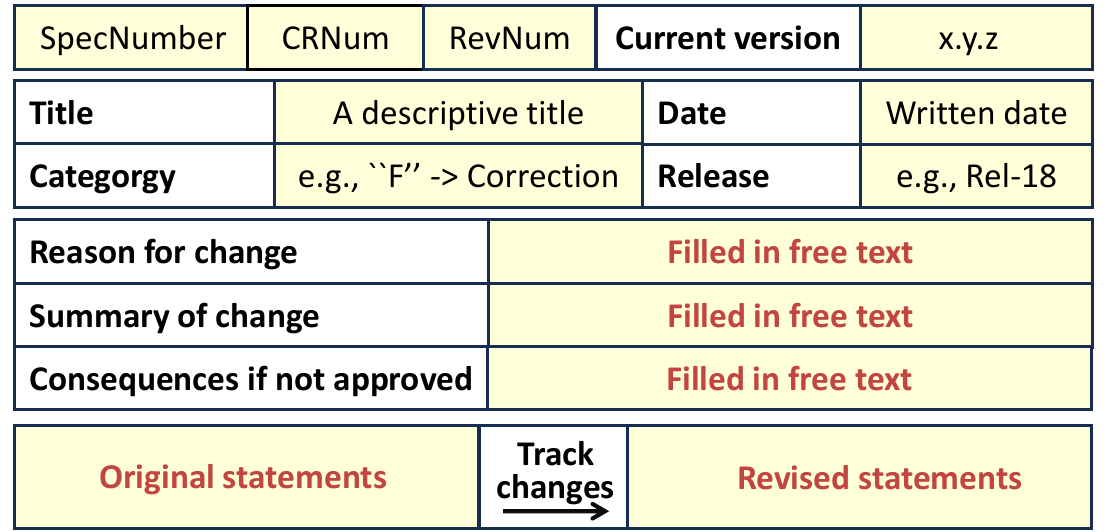}
    \caption{\label{fig:cr-illustration}Structure of 3GPP Change Request coversheet (see~\Cref{tab:cr-example-part1,tab:cr-example-part2} in~\Cref{appx:examples-and-prompts} for examples).}
    \vspace{-1em}
\end{figure}

\vspace{0.3em}
\noindent \textbf{Cellular specifications}, standardized by the 3rd Generation Partnership Project (3GPP), define the operation of cellular network systems, ensuring interoperability across vendors~\citep{gsma2025mobile-user-statistics, 3gpp-standard}.
As cellular networks evolve from 2G through 5G and beyond, specifications undergo updates through a structured process involving \textit{technical specification groups (TSGs)} and industry stakeholders.
Among these, many updates aim to address inherent security/privacy vulnerabilities discovered in cellular specifications (we refer curious readers to~\Cref{appx:spec-weaknesses} for a brief introduction).

\vspace{0.3em}
\noindent \textbf{Change request}.
To manage specification updates, 3GPP employs a Change Request (CR) procedure to revise specifications for various purposes, including keeping consistent with a change in an earlier release (\textit{A}), addition of feature (\textit{B}), functional modification of feature (\textit{C}), editorial modification (\textit{D}), and correction (\textit{F})~\citep{3gpp.21.900-cr-drafting}.
3GPP individual members (\eg, Qualcomm, Apple) raise CRs using a template coversheet~\citep{3gpp-change-requests-step-by-step}.
As illustrated in~\Cref{fig:cr-illustration}, each CR has several key blocks, including meta-information, expert rationales that explain the necessity of revisions, and the proposed clause modifications.
The modifications are tracked by the word processor software's ``revision mode'' and surrounded by the proposer-decided specification clauses as context.

\section{Benchmarking LLMs for Cellular Specification Refinement}
\label{sec:creval}

\subsection{Insight: Leveraging CRs as Data Source}\label{sec:CRs-as-domain-data}

Large-scale human labeling for the task of refining cellular specification is largely impractical due to the high demand for expertise.
To address the challenge of domain data scarcity, we propose utilizing the approved change requests as valuable data sources. 
Specifically, CRs that correct existing statements are especially suited for our purpose as they inherently reflect specification weaknesses in earlier versions.
Categories such as \textit{F} (correction), \textit{D} (editorial modification), and potentially others, encompass various specification weaknesses discussed in this work.
Key elements of our focus include \textit{expert rationales} \reasontext (\textit{reason for change} \reasonforchange and \textit{consequences if not revised} \consequencetext), \textit{summary of change} \revisionsummary, \textit{original statements} \originalstat, and \textit{revised statements} \revisedstat.
See~\Cref{tab:cr-example-part1} for a CR example.

\subsection{Formalizing Specification Refinement Into LLM-Tractable Tasks}\label{subsec:critic-tasks}

We devise three domain sub-tasks that mirror the real-world process of refining cellular specifications, as illustrated in~\Cref{fig:task-formulation}.
\begin{packeditemize}
    \item \textbf{\fillcrtitle (\fillcrsymb):} This task positions LLMs as expert reviewers, requiring them to discover potential weaknesses in given statements.  This task is relatively challenging as models receive minimal contextual information.
    \item \textbf{\outlinerevtitle (\outlinerevsymb):} Once specification weaknesses are identified, the next step is revision. To simplify the task, we require the model to outline a revision plan.
    \item \textbf{\diffanalysistitle (\diffanalysissymb):} This task supports real-world scenarios where editors assess whether revisions exactly imply and thus address the identified weaknesses.
\end{packeditemize}

Our evaluation emphasizes scenarios where LLMs operate in a zero-shot setting, where LLMs receive a general task instruction without any case-dependent inductive information (see \Cref{box:fill-cr-task-instruction} of the \fillcrtext task).
This setting honestly reflects their intrinsic ability to handle the given task instance with minimal human intervention.

\subsection{Benchmark Establishment}\label{subsec:benchmark}

Each change request can be instantiated across the three domain tasks via predefined task templates.
Concretely, each test case comprises a task instruction, a structured task-dependent input, and a reference answer.
We provide examples of the \fillcrtext task in~\Cref{example:fill-cr-1,example:fill-cr-2,example:fill-cr-3}. 
While change requests are issued for various purposes, we focus primarily on those with security-related consequences and potentially severe implications.
Using LLM-based security tagging, we select 200 security-related CRs to form \oureval.
We defer the dataset curation details to~\Cref{subsec:dataset-curation} to enable clearer side-by-side understanding of the training-set and benchmark processing (including decontamination).

We conduct a comprehensive structural analysis of the 200 test cases in~\Cref{appx:structural-analysis}:
The benchmark exhibits extensive release and specification coverage, progressive difficulty levels, and long-context complexity.
Qualitatively, the test cases in \oureval feature well-structured, focused specification clauses rich in cellular network terminology (\eg, \textit{AUTS}, \textit{VLR/SGS}, and \textit{synchronization failure message}). 
\oureval serves as a holistic assessment of LLM capabilities, encompassing extensive domain knowledge, systematic reasoning, precise instruction following, effective long-context processing, a deep understanding of cellular specification weaknesses, and acute awareness of security-related vulnerabilities. 

\begin{figure*}[t]
    \centering
    \includegraphics[width=\linewidth]{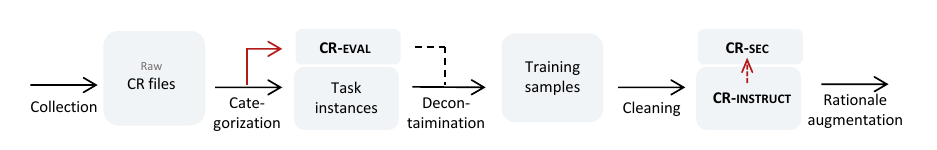}
    \vspace{-2em}
    \caption{\label{fig:data-processing-pipeline}Overview of the data-processing pipeline with change requests. This sequentially includes categorization of security relevance, decontamination, cleaning, and rationale augmentation. See~\Cref{appx:dataset-curation} for details.}
    \vspace{-1em}
\end{figure*}

\subsection{Automatic Evaluation}\label{subsec:LLM-as-a-Judge}

Following prior work~\citep{zheng2024llm-judge-fastchat,fang2024usenix-llm-for-sec-code-analysis-ref-2-LLM-as-a-Judge-sec-task-1, ullah2024llms-for-sec-code-vuln-ref-1}, we use a reference-aware, point-wise LLM-as-a-Judge setting, where each LLM-generated answer is scored by comparing it to the reference answer.
LLM-as-a-Judge evaluates responses using a 5-point Likert scale~\citep{likert1932technique}, where the positive two points indicate acceptance.
This allows differentiation between varying degrees of acceptance.
The detailed LLM-as-a-Judge prompt template is shown in~\Cref{box:evaluate-fill-cr}, with minor task-specific variations.
We instantiate the LLM-as-a-Judge with \texttt{GPT-4o}~\citep{GPT4o}.
For reproducibility, we prompt LLM-as-a-Judge to directly give back the scoring and greedily decode with temperature as 0.

We validate the reliability of our final LLM-as-a-Judge setup through a human study, involving eight PhD students majoring in network security.
The study includes two rounds: an alignment test and a judgment approval test.
Detailed settings, labeling system snapshots, and results are provided in~\Cref{appx:human-study-details}.
Key takeaways include:
\begin{packeditemize}
    \item \textbf{Availability:} Manual checking is extremely labor-intensive, underscoring the need for efficient automatic methods. LLM-as-a-Judge is rather fast and accessible.
    \item \textbf{Conformity:} 
    Human participants frequently disagree on certain LLM responses, while LLM-as-a-Judge typically yields agreements with the majority of participants.
    \item \textbf{Reliability:} Although human participants may have distinct judgment criteria, most LLM-as-a-Judge's evaluations are acceptable for them in the judgment approval test.
\end{packeditemize}

\subsection{Curating Domain-Specifc Datasets}
\label{subsec:dataset-curation}

Before diving into training methods, we introduce four domain datasets. 
We illustrate the CR processing pipeline in~\Cref{fig:data-processing-pipeline} and detail curation steps in~\Cref{appx:dataset-curation}.
These datasets include:
\begin{packeditemize}
    \item \oureval (benchmarking): 200 security-related CRs for evaluating domain-specific capabilities.
    \item \stageonedata ($\mathcal{D}_{\text{DACT}}$): A dataset for continual training on cellular specs and general knowledge.
    \item \stagetwodata ($\mathcal{D}_{\text{TST}}$): CR-converted data for fine-tuning LLMs on domain tasks.
    \item \stagethreedata ($\mathcal{D}_{\text{SCT}}$): Security-related CR data for enhancing LLMs' focus on security weaknesses.
\end{packeditemize}

\section{Domain Specialization via Fine-Tuning}
\label{sec:model-training}

In this section, we resort to fine-tuning to achieve domain specialization.
We have also explored prompting-based methods in~\Cref{appx:prompt-limit}.

\begin{figure*}[t]
    \centering
    \includegraphics[width=\linewidth]{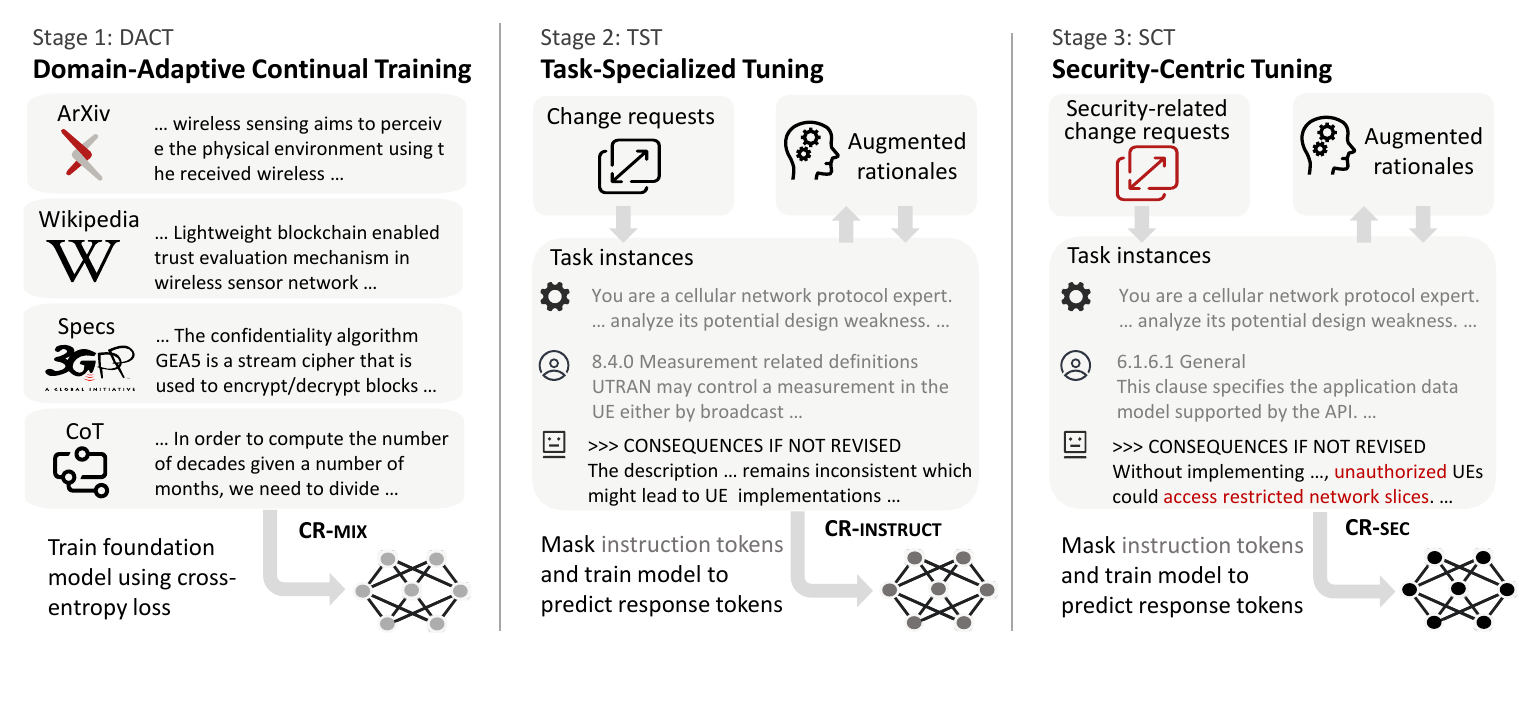}
    \vspace{-3em}
    \caption{\label{fig:overview}High-level overview of the training framework for domain specialization.}
    \vspace{-1em}
\end{figure*}

In the following, we propose a three-stage training framework that mirrors human expert development, as illustrated in~\Cref{fig:overview}.
We also propose rationale augmentation in~\Cref{subsec:rationale-augmentation} for converting raw CRs into high-quality training data.

\subsection{Three-Stage Training Framework}

\noindent \textbf{Stage 1: Domain-Adaptive Continual Training (DACT).}
As we cannot assume that foundation models have adequately acquired domain knowledge during their initial pre-training, we refine an LLM's learned distribution through continual pre-training on domain data~\citep{gururangan2020donot-stop-pretraining-dapt-concept, zhou2024lima,ghosh2024closer-look-at-ift-limitations}. 
This is modeled as follows:
\begin{equation}
\mathcal{L}_{\text{DACT}}(\theta) = -\mathbb{E}_{x \sim \mathcal{D}_{\text{DACT}}} \left[\sum_{i=1}^{n} \log p_\theta(x_i|x_{<i})\right],
\end{equation}
where we instantiate $\mathcal{D}_{\text{DACT}}$ with \stageonedata.

\noindent \textbf{Stage 2: Task-Specialized Tuning (TST).}
This stage is designed to help LLM master the basic ability to analyze cellular specifications.
Fine-tuning during this stage utilizes our \stagetwodata dataset, which encompasses all CR data, and relies on labeled samples:
\begin{equation}
\mathcal{L}_{\text{TST}}(\theta) = -\mathbb{E}_{(x,y) \sim \mathcal{D}_{\text{TST}}}
\left[ \log p_{\theta}(y \mid x) \right],
\end{equation}
where $\mathcal{D}_{\text{TST}}$ represents the \stagetwodata dataset.
The \stagetwodata dataset incorporates diverse task formulations, enabling the model to learn each CR through multiple contexts.
This multi-task learning paradigm encourages the model to generalize reasoning skills across tasks.

\noindent \textbf{Stage 3: Security-Centric Tuning (SCT).}
To tackle \oureval, we expect the model to analyze specifications from the security perspective.
Inspired by~\citet{he2023sven-ccs-llm-for-sec-code-ref}, we frame security-centric analysis as a style-controlled text generation problem.
This approach leverages security-related CRs, which reveal real-world security issues, to shape security-centric analysis, enhancing operational feasibility.
The loss is defined as:
\begin{equation}
\mathcal{L}_{\text{SCT}}(\delta\theta) = -\mathbb{E}_{(x,y) \sim \mathcal{D}_{\text{SCT}}} \left[\log p_{\theta + \delta\theta}(y \mid x) \right],
\end{equation}
where $\mathcal{D}_{\text{SCT}}$ denotes security-related task instances of the target task from \stagethreedata.
The parameter $\delta\theta$ corresponds to the additional adapter implemented using LoRA~\citep{hu2022lora}, which allows for efficient adaptation and preserves most of the model's original capabilities~\citep{biderman2024lora-learn-less-and-forget-less}.

\subsection{Rationale Augmentation}\label{subsec:rationale-augmentation}
Manual inspection reveals a limitation of training samples derived from CRs: rationales \reasontext written by human experts in CRs are mainly concise declarative statements, rather than detailed reasoning (see Appendix~\Cref {fig:rationale-augmentation-example} for an example).
This gap hinders effective LLM training, as insufficient rationales lead models to memorize answers instead of developing problem-solving skills~\citep{chung2024scaling-instruction-fine-tuned-llm, kim2023cot-collection-dataset-rationalization-ref-3, yue2024mammoth}.

To address this issue, we introduce rationale augmentation, generating refined, rationale-rich responses for LLM training.  
Following prior work on training with rationales~\citep{rajani2019explain-yourself-rationalization-ref--2, zelikman2022star, kim2023cot-collection-dataset-rationalization-ref-3}, we adopt a backward-rationalization strategy:  
A rationale generator \( P_{\text{o}} \) processes a complete task instance—comprising task instruction \( T \), test case \( Q \), and original answer \( A \)---and applies backward reasoning to produce a rationale-augmented answer \( A^* \), following augmentation principles \( C \), as formulated by \( A^* \leftarrow P_{\text{o}}(C \mid T \oplus Q \oplus A) \). 
We enforce pedagogically oriented principles $C$ to enhance instructional effectiveness while preserving answer consistency (see~\Cref{box:enrich-answer-rationale}).

\definecolor{DeltaBlue}{HTML}{4f86ec}
\definecolor{rowblue}{HTML}{ECF4FC}

\definecolor{highlightred}{HTML}{c00000}
\newcommand{\bestcell}[1]{\textcolor{highlightred}{\textbf{#1}}}

\setlength{\fboxrule}{0.6pt}

\begin{table*}[t!]
\centering
\begin{minipage}{\textwidth}
    \caption{LLMs' performance in \oureval across the three domain tasks. We highlight the best performance within open-source and closed-source LLMs, respectively.}
    \label{tab:benchmark-results}
    \renewcommand{\arraystretch}{1.06}
    {
    \small
    \setlength\tabcolsep{5pt}
    \resizebox{0.97\textwidth}{!}{%
    \begin{tabular}{p{0.21\textwidth} ccc ccc ccc}
    \mytoprule{1-10}
    & \multicolumn{3}{c}{\textbf{\fillcrtitle}} & \multicolumn{3}{c}{\textbf{\outlinerevtitle}} & \multicolumn{3}{c}{\textbf{\diffanalysistitle}}  \\
    \multicolumn{1}{c}{\textbf{\large Model}} & \multicolumn{3}{c}{\textbf{\fillcrsymb}}& \multicolumn{3}{c}{\textbf{\outlinerevsymb}} & \multicolumn{3}{c}{\textbf{\diffanalysissymb}}  \\
    \cmidrule(lr){2-4}  \cmidrule(lr){5-7}  \cmidrule(lr){8-10} 
    &  \passatk{1} & \passatk{3} & \passatk{5} & \passatk{1} & \passatk{3} & \passatk{5} & \passatk{1} & \passatk{3} & \passatk{5} \\     
    \mymidrule{1-10}     \rowcolor[gray]{0.9}
    \multicolumn{10}{c}{\textbf{\textit{Open-Source LLMs}}}\\
    \mymidrule{1-10}
    GLM-4-9B  & 14.5 & 23.3 & 27.8 & 172.7 & 188.3 & 191.0 & 28.7 & 50.1 & 61.3  \\
    GLM-4.5  & 16.0 & 23.0 & 26.4 & 180.2 & 186.6 & 188.0 & 101.6 & 117.6 & 122.4 \\
    Mistral-7B-v0.3 & 9.1 & 16.1 & 19.8 & 163.4 & 182.5 & 186.5 & 26.0 & 44.9 & 54.9  \\
    InternLM-2.5-7B & 12.9 & 25.9 & 33.5 & 158.9 & 185.3 & 190.8 & 21.9 & 42.4 & 54.7  \\
    Qwen-2.5-7B & 13.9 & 24.9 & 29.9 & 175.5 & {189.3} & {191.6} & 32.0 & 55.7 & 68.2  \\
    Qwen-2.5-14B & 17.6 & 27.4 & 30.8 & 183.6 & 193.9 & 196.0 & 85.8 & 119.1 & 130.1  \\
    Qwen-2.5-32B & 18.0 & 28.8 & 33.7 & 183.2 & 190.6 & 192.6 & 77.2 & 106.7 & 116.6   \\
    Qwen-2.5-72B  & 15.2 & 22.4 & 25.7 & 186.2 & {195.5}  & {197.7} & 79.4 & 105.5 & 114.1 \\
    Qwen3-8B & 15.4 & 25.8  & 30.8 & 188.1 & 195.6 & 197.2 & 58.0 & 85.4 & 95.6 \\
    Qwen3-14B & 23.3 & 37.6 & 44.5 & 185.1 & 194.4 & 196.5 & 81.3 & 112.5 & 122.7 \\
    Qwen3-32B & 21.5 & 35.6 & 42.1 & 188.0 & 195.4 & 197.2 & 96.1 & 130.4 & 142.5 \\
    Qwen3-30B-A3B & 18.5 & 31.8 & 38.2 & 188.1 & 194.2 & 195.2 & 72.8 & 102.3 & 114.6 \\
    Qwen3-235B-A22B & 23.5 & 36.4 & 41.9 & \bestcell{192.9} & 196.7 & 197.6 & 111.1 & \bestcell{144.2} & \bestcell{155.1} \\
    DeepSeek-V3 & 8.4 & 13.8 & 16.7 & {188.4} & 195.1 & 197.0 & 95.6 & 121.5 & 128.8  \\
    DeepSeek-R1 & 9.2 & 15.8 & 19.4 & {192.0} & \bestcell{197.2} & \bestcell{198.3} & \bestcell{119.2} & {143.7} & 151.3 \\
    LLaMA-3.1-70B  & 7.4 & 13.4 & 16.4 & 144.4 & 168.9 & 174.8 & 40.5 & 64.3 & 76.1 \\
    \hdashline
    LLaMA-3.1-8B & 6.1 & 13.2 & 18.1 & 126.4 & 164.2 & 174.0 & 27.4 & 48.3 & 59.8  \\
    \rowcolor{rowblue}
    \textbf{\textsc{CRitic}}-LLaMA-3.1-8B  & \bestcell{{27.2}} & \bestcell{{42.3}} & \bestcell{57.8} & 160.5 & 182.4 & 186.7 & {106.4} & {137.9} & {148.4} \\
\rowcolor{rowblue}
\textit{$\boldsymbol{\Delta}$ (Relative Increase) }
& \textcolor{DeltaBlue}{\textbf{+345.9\%}}
& \textcolor{DeltaBlue}{\textbf{+220.5\%}}
& \textcolor{DeltaBlue}{\textbf{+219.3\%}}
& \textcolor{DeltaBlue}{\textbf{+27.0\%}}
& \textcolor{DeltaBlue}{\textbf{+11.1\%}}
& \textcolor{DeltaBlue}{\textbf{+7.3\%}}
& \textcolor{DeltaBlue}{\textbf{+288.3\%}}
& \textcolor{DeltaBlue}{\textbf{+185.5\%}}
& \textcolor{DeltaBlue}{\textbf{+148.2\%}} \\
    \mymidrule{1-10}
     \rowcolor[gray]{0.9}
    \multicolumn{10}{c}{\textbf{\textit{Closed-Source LLMs}}}\\
    \mymidrule{1-10}
    Doubao-seed-1-6-flash  & 21.3 & 33.3 & 39.1 & 159.8 & 173.9 & 178.0 & 88.5 & 117.6 & 127.9 \\
    Doubao-seed-1-6-lite  & 18.3 & 28.9 & 33.7 & 171.4 & 182.0 & 184.6 & 124.4 & 153.5 & 161.3 \\
    Doubao-seed-1-6 & 24.6 & 41.2 & 48.4 & 190.7 & 197.5 & 198.5 & 144.4 & 169.5 & 175.1 \\
    Doubao-seed-1-6-thinking  & 42.0 & 63.0 & 72.9 & 189.7 & 196.0 & 197.1 & 158.9 & 177.7 & 182.2 \\
    Claude-Sonnet-3.5 & 9.5 & 16.2 & 19.3 & 172.6 & 182.5 & 184.9 & 77.7 & 106.3 & 118.1 \\
    Gemini-2.0-flash-thinking & 79.0 & 114.8 & 127.3 & 166.8 & 177.6 & 179.2 & {139.8} & {164.0} & {169.5} \\
    Gemini-2.5-flash & 61.8 & 92.1 & 106.4 & 181.4 & 192.0 & 194.7 & 158.1 & 178.0 & 182.6 \\
    Gemini-2.5-pro & 62.7 & 93.0 & 106.0 & 185.3 & 195.6 & 197.8 & 172.7 & 184.8 & 187.2 \\
    GPT-3.5-turbo & 11.2 & 20.1 & 24.2 & 146.2 & 166.3 & 170.7 & 42.2 & 63.6 & 71.7 \\
    GPT-4o-mini & 18.2 & 27.5 & 31.2 & 173.0 & 182.3 & 183.5 & 52.4 & 74.0 & 81.8 \\
    GPT-4o & 16.0 & 25.3 & 29.2 & {176.8} & 186.3 & 188.0 & 88.0 & 113.5 & 122.6 \\
    GPT-o3-mini & \bestcell{89.0} & \bestcell{116.6} & \bestcell{127.9} & 186.8 & 192.5 & 194.0 & 132.5 & 154.4 & 162.0 \\
    GPT-5-mini & 73.9 & 105.4 & 118.9 & \bestcell{199.7} & \bestcell{200.0} & \bestcell{200.0} & 171.6 & 181.0 & 183.7 \\
    GPT-5 & 75.3 & 109.8 & 123.9 & 196.3 & 199.0 & 199.7 & \bestcell{182.7} & \bestcell{190.2} & \bestcell{192.6} \\
    \mybottomrule{1-10}
    \end{tabular}%
    }
    }
\end{minipage}
\end{table*}

\section{Experiments}

\subsection{Experimental Setup}\label{subsec:exp-setup}

\noindent \textbf{Metrics.} 
We evaluate LLM performance on \oureval using \passk~\citep{chen2021oai-codex-and-humaneval}. 
The \passk metric measures the success rate by allowing $k$ independent attempts and considering the best result among the $k$ completions.
Given $n \geq k$ completions, where $c \leq n$ completions are correct (\ie, accepted by the LLM-as-a-Judge), the unbiased \passk score is computed as:
$\text{pass@}k := 1 - \frac{\binom{n - c}{k}}{\binom{n}{k}}$
Specifically, we report the cumulative \passk score over all test cases, with a maximum of 200.
Following established practices~\citep{roziere2023code-llama, chen2021oai-codex-and-humaneval, gu2024cruxeval-800samples-input-prediction-and-output-prediction}, we set the sampling temperature to 0.8 and top-p to 0.95.
Balancing reliability and cost, we sample $n=10$ completions.

\noindent \textbf{Models.}
Our evaluation captures the currently highest achievable performance of LLMs, including 17 open-source models and 14 closed-source models.
For open-source models, we use the official chat templates.
For reasoning models, we set the reasoning efforts to medium.
\Cref{tab:evaluated-models} details the models.

\noindent \textbf{Rationale augmentation.}
In this work, we employ \texttt{LLaMA-3.1-70B} and \texttt{GPT-4o} for rationale augmentation of \stagetwodata and \stagethreedata, respectively.
We prompt the rationale generators with a temperature of 0.8 and a top-p of 0.95 to encourage rationale diversity.
Our default rationale number per instance is three for TST and five for SCT.
Note that we do not augment the reference answers on \oureval, avoiding biased evaluation.

\noindent \textbf{Training configurations.}
Constrained by resource limit, our experiments are primarily on {\raggedright \texttt{LLaMA-3.1-8B}}~\citep{dubey2024llama-3-1-meta-llm-advance-ref-3}, a representative 8B model at the inception of the project.
Meanwhile, we acknowledge and have verified that fine-tuning on stronger base models can achieve better domain abilities, as explored in~\Cref{appx:sft-other-models}.
We fine-tune all parameters for the first two stages, DACT and TST.
Then, we introduce different LoRA adapters~\citep{hu2022lora} ($r=256$, $\alpha=512$) for the three domain tasks in SCT.
The training ends up with \ourmodel\footnote{We name it \critic, for the model is trained to act as a \textbf{critic} for cellular network specifications, and its power can be attributed to \textbf{C}hange \textbf{R}equests.}.
We provide a more detailed list of our training choices in~\Cref{appx:training-configs}.

\subsection{Evaluation Results}\label{subsec:evaluation-results}

\noindent \textbf{General performance.}
As shown in~\Cref{tab:benchmark-results}, the three tasks vary in difficulty, from the easiest \outlinerevtext to the hardest \fillcrtext.
The \outlinerevtext task, primarily a summarization task, is well-handled by most models, with some smaller models (\texttt{Qwen-2.5-7B}, \texttt{GLM-4-9B}) even outperforming closed-source counterparts (\eg, \texttt{GPT-4o}).
In contrast, the \diffanalysistext task reveals a significant gap between models, highlighting the challenge of identifying implicit specification weaknesses even when given revisions as hints.
The \fillcrtext task emerges as an extremely challenging task, particularly for open-source models.
Proprietary reasoning models (\texttt{GPT-o3-mini}, \texttt{Gemini-2.0-flash-thinking}) demonstrate superior performance, particularly in the \diffanalysistext and \fillcrtext tasks. 
This suggests that reasoning may be a key enabler of strong analysis of sophisticated specifications, which is also emphasized by our rationale augmentation design.

\noindent
\textbf{Gap between general and domain-specific capabilities.}
Several widely recognized LLMs, such as \texttt{DeepSeek-R1} and \texttt{Claude-3.5-Sonnet}, perform poorly on \fillcrtext. 
We also observe inverse scaling~\citep{mckenzie2023inverse-scaling} in the \texttt{Qwen-2.5} family, where the large 72B model performs the worst in the \fillcrtext task compared to its smaller-sized cousins. 
These underscores \textit{a potential gap between general-purpose LLMs and domain-specific task requirements} while emphasizing the importance of \oureval in helping practitioners identify models with the strongest domain-specific capabilities.

\begin{figure}[!t]
    \centering
    \includegraphics[width=\linewidth]{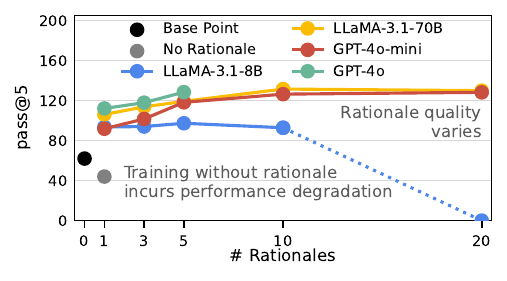}
    \vspace{-2.5em}
    \caption{\label{fig:exp-rationale-scaling}Rationale-dimension scalability.}
    \vspace{-1em}
\end{figure}

\subsection{Performance Analysis of Fine-Tuning}\label{subsubsec:scalability}

\noindent \textbf{Performance of the domain-specialized LLM.}
\ourmodel nearly triples \passatk{5} scores of its base model in the \diffanalysistext (+219.3\%) and \fillcrtext (+148.2\%) tasks. 
Notably, it outperforms its \textbf{contemporary} proprietary models like \texttt{GPT-4o}, solving almost twice as many \fillcrtext test cases.
Despite these advancements, we have to acknowledge that state-of-the-art LLMs that are published more recently, particularly the reasoning models, continue to improve at an unprecedented pace.
At the time of writing, closed-source reasoning models like \texttt{GPT-o3-mini} have surpassed our domain-specialized medium-sized model, but the fine-tuned model still surpasses open-source reasoning models in \fillcrtext.
Overall, these results highlight the impact of domain-specific fine-tuning in bridging the gap between general LLM capabilities and domain-specific requirements.

\noindent \textbf{Impact and scalability of rationale augmentation.}
Rationale augmentation offers an additional scaling axis to enhance performance.
We evaluate this on \stagethreedata in the \diffanalysistext task with various rationale generators.
The landscape of scaling up rationales is illustrated in~\Cref{fig:exp-rationale-scaling}.
\textit{Training without rationales degrades performance, confirming that raw task instances offer limited learning value,} underscoring the necessity of rationale augmentation.
Typically, the benefits of incorporating more rationales show evident gains and then reach a plateau.
More capable rationale generators yield greater performance gains, likely due to improved knowledge distillation~\citep{gou2021knowledge-distillation-survey}.
An exception is observed with \texttt{LLaMA-3.1-8B}, where training with highly diverse rationales (\eg, 20) generated by itself leads to model collapse (see~\Cref{appx:explain-rationale} for a preliminary explanation).

\begin{table}[t]
\centering
\caption{Analysis of token prediction behavior in \ourmodel on \oureval. Note: '\_' denotes the blank character in tokens. A more detailed version is provided in~\Cref{tab:token-behavior-full}.}
\label{tab:token-behavior-selected}
\Large
\resizebox{0.47\textwidth}{!}{
\begin{tabular}{cccc}
\toprule
\textbf{Token} & \textbf{Ratio} & \textbf{Token} & \textbf{Ratio} \\
\midrule
       \_safeguard & $138.70 \times$ & \_degrade & $79.21 \times$ \\
       \_improper & $58.78 \times$ & Failure & $61.47 \times$ \\
       \_mistakenly & $39.49 \times$ & \_challenges & $47.90 \times$ \\
       \_interception & $32.64 \times$ & \_interruptions & $24.07 \times$ \\
       \_inadvertently & $21.85 \times$ & \_reuse & $16.60 \times$ \\
       \_misuse & $13.98 \times$ & \_operational & $16.38 \times$ \\
       \_susceptible & $12.34 \times$ & \_cryptographic & $10.98 \times$ \\
       \_legal & $10.49 \times$ & \_degraded & $10.07 \times$ \\
\bottomrule
\end{tabular}
}
\end{table}

\subsection{Analyzing Model Behaviors}\label{subsec:explain-effectiveness}

We investigate why \ourmodel excels on \oureval by analyzing its next-token prediction behavior on the \diffanalysistext task.
For each test case, we collect softmax-normalized next-token distributions and apply hierarchical aggregation: averaging predictions across tokens within each sample, then across all samples.
See~\Cref{appx:track-behavior} for more operational details.
This yields a single distribution $P_{\text{LLM}}\in\mathbb{R}^{|V|}$ that summarizes the model’s overall token-level behavior on \oureval.
We compare \ourmodel with its base model, \text{LLaMA-3.1-8B}, highlighting representative differences in~\Cref{tab:token-behavior-selected}.
\ourmodel assigns higher probabilities to security-related tokens (\eg, ``\_safeguard'') and shifts toward more specific terminology (\eg, ``\_interception''), reflecting increased domain specialization.

\paragraph{\ding{117}\ Additional experiments.}
In~\Cref{appx:additional-exps}, we conduct more experiments to (1) ablate the three training stages, (2) scale training data, (3) fine-tune additional backbones (\eg, \texttt{GPT-4o-mini}), and (4) study transfer across \oureval\ tasks.

\section{Case Study: Can LLMs Replace Humans in Discovering Weaknesses?}\label{subsec:known-vulns}

To understand this, we test LLMs in discovering 3G/4G/5G vulnerabilities that were previously reported in top-tier security conferences.

\noindent \textbf{Experimental setup}.
Following~\citet{al2024hermes}, we examine a total of 36 types of existing attacks (see~\Cref{tab:known-vulns}), collecting relevant specification statements, threat models, and attack descriptions for each.
This process is labor-intensive, costing around five working days of two students with extensive experience in cellular security.
Specification statements were successfully located for 30 attacks, while the remaining were excluded due to reliance on additional implementation or configuration flaws.
The examination is conducted as follows:
1) \ourmodel (\fillcrtext) analyzes potential weaknesses across 10 trials.
2) \texttt{GPT-4o} evaluates each analysis according to~\Cref{box:gpt-verifier}, determining whether combining the discovered weakness with the corresponding threat model sufficiently derives the final attack.
3) We manually validate accepted analyses for reliability.

\begin{table}[t]
\centering
\small
\caption{\label{tab:known-vulns-examples}Examples of \ourmodel analyses for the \textit{Incarceration with RRCRELEASE attack}~\citep{hussain20195greasoner}.}
\renewcommand{\arraystretch}{1.07}
\begin{tabular}{p{0.47\textwidth}}
\toprule
\textbf{Accepted \ourmodel response} \\ \hline
The current specification allows the RRCRelease message to be sent unprotected under certain conditions. This poses a significant security risk, particularly if these messages contain sensitive information or instructions that \textcolor{takeawaycolor}{could be intercepted or modified by malicious entities}. The inclusion of optional fields such as de-prioritisation information and cell reselection priorities in the RRCRelease message further exacerbates this risk, as these fields can convey critical operational data that, if leaked, could disrupt network performance or lead to unauthorized access. \\
\midrule
\textbf{Rejected \ourmodel response} \\ \hline
The current specification allows the RRCRelease message to be sent unprotected, which can lead to significant security vulnerabilities. Since the RRCRelease message may contain sensitive information, such as redirected carrier information or suspend configuration details, \textcolor{takeawaycolor}{sending it unprotected could allow unauthorized entities to intercept this data}. \\
\bottomrule
\end{tabular}
\end{table}

\noindent \textbf{Quantitative results}.
\ourmodel detects all 30 attack types.
In contrast, the conventional formal analysis method Hermes~\citep{al2024hermes} detects 19.
Due to space constraints, full results are reported in Appendix~\Cref{tab:known-vulns}. 

\noindent \textbf{Qualitative examples}.
Representative examples of \ourmodel's responses are shown in~\Cref{tab:known-vulns-examples}.
The domain-specialized model exhibits the ability to reason about potential weaknesses diversely, which we attribute to the incorporation of multiple rationales during training.
Remarkably, even the rejected responses provide valuable insights, unveiling other negative consequences with the unprotected \textit{RRCRelease} message.

\noindent \textbf{Failure analysis}.
While these results are promising, we also identify several challenges, as discussed in~\Cref{appx:future-works}, \eg, low calibration due to potential hallucination~\citep{zhang2023siren-song-survey-hallucination-llms} and requirements for additional preparations.

\section{Related Works}
\label{sec:related-works}

\noindent \textbf{LLMs for cellular network}.
The human-like intelligence of modern LLMs has catalyzed numerous studies on their potential applications in cellular networks.
Existing research predominantly investigates whether LLMs can comprehend domain knowledge through question-answering tasks.
For example, GSMA has officially launched the Open-Telco LLM Benchmark project~\citep{gsma2025open-telco-llm-benchmarks} to evaluate LLMs on interacting with complex standards.
Similar efforts include SPEC5G~\citep{karim2023spec5g}, TSpec-LLM~\citep{nikbakht2024tspecllmopensourcedatasetllm}, and TeleQnA~\citep{maatouk2023teleqna}.
Beyond knowledge comprehension, \citet{wen20246g-hotnet-llm-explain-data-traffic-ref-1} employ LLMs to detect and explain runtime anomalies in the O-RAN data plane, while \citet{kotaru2023adapting-llm-for-operator-data-analysis} investigate their potential for analyzing 5G operator network data.
In this work, we stress-test LLMs in a productive setting, evaluating LLMs in refining cellular specifications.

\noindent \textbf{NLP for analyzing cellular network specifications}.
NLP techniques have been adopted to uncover specification flaws~\citep{chen2021bookworm, chen2022creek, chen2023sherlock, rahman2024cellularlint, al2024hermes}.
Atomic~\citep{chen2021bookworm} applies textual entailment to detect risky descriptions in 3GPP standards.
Several approaches fine-tune encoder models (\eg, RoBERTa~\citep{liu2019roberta}) for different purposes:
CREEK~\citep{chen2022creek} identifies security-related CRs, CellularLint~\citep{rahman2024cellularlint} detects inconsistencies, and Hermes~\citep{al2024hermes} constructs state machines for formal analysis.
These methods focus on information extraction rather than direct vulnerability detection.

\section{Conclusion}\label{sec:conclusion}

In this work, we pioneer the adoption of LLMs for automated cellular specification refinement.
To advance it, we tackle the domain data scarcity challenge by transforming change requests of 3GPP standards into utilizable task instances and formulate three domain tasks.
We establish the \oureval benchmark, enabling the community to assess the domain-specific capabilities of the rapidly advancing LLMs.
What's more, we enhance LLM domain specialization by contributing effective training recipes.
Our case studies on 30 known cellular attacks reveal the current status in achieving fully automated cellular specification refinement.
This study sheds light on the potential of LLMs and provides a foundation for future advancements.

\clearpage

\section{Limitations}

\noindent \textbf{Coverage of models}.
Due to cost constraints, our measurement study does not cover all existing LLMs; instead, we focus on recent state-of-the-art models as representative examples.
Future work may include newly released LLMs.
For fine-tuning, we mainly focus on LLaMA-3.1-8B, and also experiment with LLaMA-3.1-70B and GPT-4o-mini.
These models are all chat-style models.
Although we do not explicitly train reasoning models, we hypothesize that they may achieve better performance after domain specialization.

\noindent \textbf{Absence of human baselines}.
As pinpointed in~\Cref{appx:structural-analysis}, \oureval involves test cases that are related to as many as 74 distinct specifications.
Among these, the longest specification can span more than 1,600 PDF pages (\eg, 3GPP TS 38.331 v18.0.2 for 5G RRC).
In fact, these specifications are typically studied and updated by different experts or editors.
This renders it challenging to recruit experts who can grasp all the involved specifications.
However, as test cases are converted from expert-issued change requests, the ground truths can be seen as an ensemble of human performance.

\noindent \textbf{Full completeness of test instances}.
In this work, we directly convert change requests into test instances.
This means that the provided context for LLMs only contains those specification clauses that are directly related to the weakness and thus cannot ensure guarantees of self-inclusiveness.
Actually, this is a deliberate choice:
As mentioned in~\Cref{subsec:benchmark}, the evaluated capabilities by \oureval are not simply limited to the reasoning ability for specification analysis, but extend to domain knowledge about cellular specifications.

\noindent \textbf{Optimal fine-Tuning choices}.
Our systematic exploration consumed over 32,120 H800 GPU hours, but computational constraints prevented us from exploring other promising directions, \eg, scaling to giant LLMs.
Nevertheless, our experiments in~\Cref{subsubsec:scalability} demonstrate the feasibility of extending domain specialization to stronger base models.

\noindent \textbf{Alternative domain specialization techniques}.
This work mainly explores fine-tuning and prompting to enhance LLMs' domain-specific capabilities. 
Alternative approaches such as reinforcement learning and agentic AI may further boost domain specialization, which we leave for future work.

\noindent \textbf{Scope of weakness types}.
Additionally, while our focus is on the security aspects of cellular specifications, our future work could extend LLM-driven specification refinement to address other types of weaknesses following similar methodologies. 
Meanwhile, this work mainly focuses on cellular specifications, which is a subject of crucial importance.
However, our exploration can also be extended to other protocols, \eg, IoT protocol, DNS protocol, or even proprietary protocols.

\noindent \textbf{Cross-lingual evaluation}.
In this work, we mainly focus on evaluating LLMs in English, as the 3GPP committee uses English as the working language for both specifications and change requests most of the time.
An interesting future work is to investigate the domain-specific performance of LLMs in a cross-lingual setting.

\section{Ethical Considerations}

Our research faithfully adheres to the ethical guidelines established by the Association for Computational Linguistics (ACL)\footnote{\url{https://aclrollingreview.org/responsibleNLPresearch/}}. 
Our use of AI assistants in this manuscript is limited to writing polishing under full human monitoring.

All change requests and specifications are openly accessible, which we use simply for research purposes.
All experiments were performed on publicly available models or API services, ensuring compliance with relevant terms of service.

In this work, our evaluation primarily focuses on existing CRs and known specification weaknesses, both of which have been previously disclosed to relevant stakeholders (e.g., 3GPP councils) and are open to the public.
We conducted all experiments and explorations in isolated environments, without allowing autonomy to exploit the weaknesses.
Our human study for the LLM-as-a-Judge validity does not contain offensive content, resulting in no harm to the annotators.

\noindent \textbf{Broader implications}.
This research aims to understand the possibility of applying large language models for automated cellular network specification refinement, which is beneficial to relevant stakeholders and practitioners.
We also recognize that the advancement of AI techniques presents a double-edged sword: while offering noteworthy benefits, they may also pose unprecedented threats to human society~\citep{bengio2024managingairisk-science}.
We should also envision the possible situation where AI systems are given more autonomy.
In this context, our work also serves as a systematic and rigorous assessment of the dangerous red-line capabilities of automated AI systems, specifically discovering and leveraging vulnerabilities within cellular specifications.
From another aspect, the application of LLMs for automated cellular specification refinement is not to replace relevant stakeholders but to provide assistance for human experts.

We will release our codebase and benchmark to ensure reproducibility of our results. 
Meanwhile, it is worth noting that transitioning from a small-scale study to a tool that can be used in the real world requires additional research to ensure the safety, reliability, and efficacy of the technology.
Finally, we remind the readers that any techniques introduced in this paper should be applied ethically and within appropriate research contexts.

\bibliography{refs}

\clearpage
\appendix
\crefalias{section}{appendix}
\crefalias{subsection}{appendix}

\section{Future Works}
\label{appx:future-works}

\begin{packeditemize}
\item {\textbf{Requirements for effective calibration:}}
While LLMs’ ability to produce diverse interpretations can be useful, it inevitably increases false positives. 
Their inherent hallucination issues~\citep{zhang2023siren-song-survey-hallucination-llms} further exacerbate this, making blind reliance infeasible.
For weakness verification, current practices primarily delegate this responsibility to human analysts~\citep{chen2021bookworm, hussain2018lteinspector, hussain20195greasoner}.
Extending this practice to LLM-based analysis would be impractical due to the sheer volume of generated weaknesses.
Crucially, we argue that even with full autonomy, decision-making should not be ceded to LLMs.
Rather than blaming LLMs, the focus should be on developing effective calibration mechanisms to reduce human effort.
Without them, unverified proposals risk overwhelming analysts instead of aiding them.

\item {\textbf{Completeness of analyzed clauses:}} 
Our manually curated set of attack-related specification clauses provides an idealized benchmark, containing sufficient context for analysis.
This partially explains why \ourmodel achieves perfect detection of known attacks despite its limitations in addressing all test cases in \oureval.
However, practical challenges emerge when refining active specifications, particularly in identifying vulnerabilities arising from complex interactions across multiple sources.
\end{packeditemize}

\section{Exploring Prompting Methods}\label{appx:prompt-limit}

\subsection{Why Fine-Tuning?}\label{appx:why-fine-tuning}
In this work, we explore fine-tuning methods to improve LLMs in cellular specification refinement.
Several compelling reasons motivate our resorting to fine-tuning:
\ding{202} Refining cellular specifications is reasoning-heavy, knowledge-intensive, and expertise-driven.
There exists a gap between LLMs' general-purpose training objectives and the task's specialized requirements.
\ding{203} As we cannot assume that general-purpose LLMs inherently possess all the fundamental components necessary for expert-level analysis, we demand an approach that enables building these capabilities from the ground up.
\ding{204} The cellular security community continuously evolves with ongoing research and new discoveries. 
This necessitates a scalable approach capable of knowledge ingestion.
Fine-tuning effectively fulfills all the expectations outlined above~\citep{chung2024scaling-instruction-fine-tuned-llm, longpre2023flan-collection-2022-teasing-apart-instruction-tuning-designs, wei2022flan-zero-shot-learners}.
\ding{205} Furthermore, for domain-specific tasks, fine-tuned models may achieve either higher performance at fixed cost or lower cost at fixed performance.
These benefits align with the success of specialized LLMs in various fields, including medicine~\citep{singhal2023large-llm-for-medicine-ref-nature}, finance~\citep{wu2023bloomberggpt-llm-for-finance-ref}, and math~\citep{azerbayev2024llemma}.

\subsection{Prompting Methods}

Another potential method to achieve domain specialization is prompting, which involves crafting well-suited instructions to steer general-purpose LLMs.
Prompting is lightweight, as it requires no additional model training.
However, it also suffers from limitations, including model capacity bottleneck, lack of systematic methodology, reliance on human expertise, limited performance scalability, restricted transferability across models, and sometimes practical policy constraints.

\begin{figure}[t]
    \centering
    \includegraphics[width=1\linewidth]{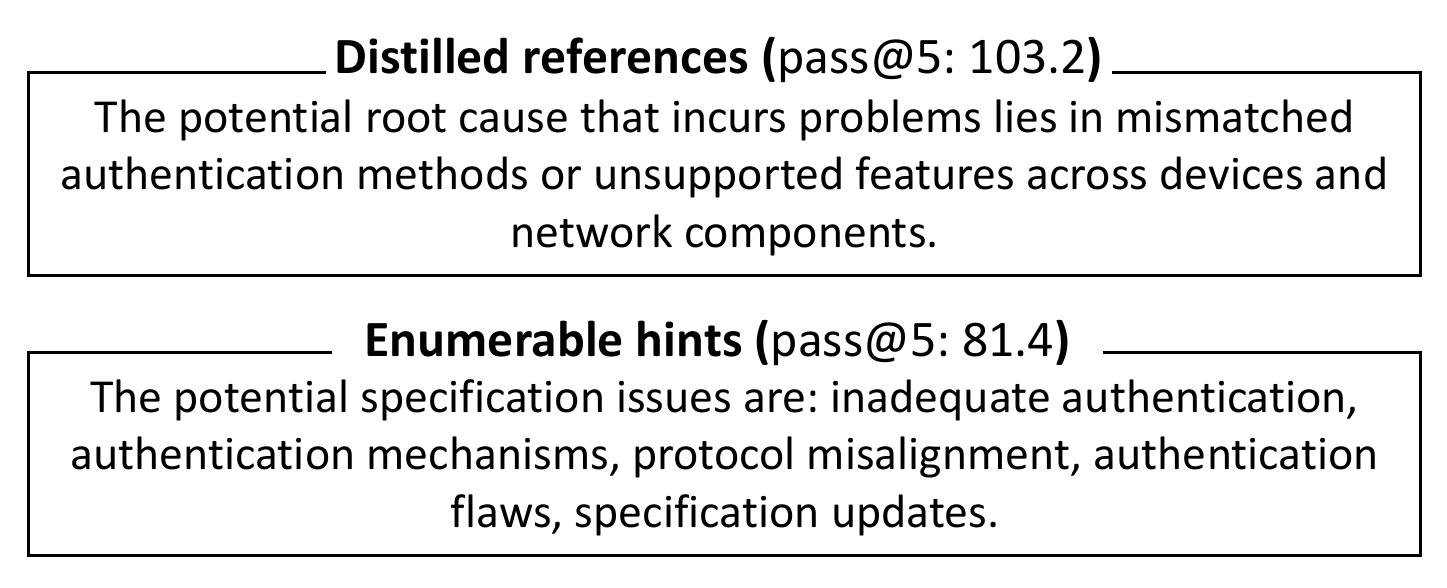}
    \caption{Results and examples of incorporating expertise. We test on the \fillcrtext task, and the \passatk{5} of the tested checkpoint with no hint is 57.8.}

    \label{fig:augment-with-hints}
\end{figure}

We conduct experiments to assess the effectiveness of various prompting methods on \oureval.
Due to space limit, we present details in~\Cref{appx:prompt-limit} and summarize the main results here:
(1) We ask one author to manually rephrase instructions or query \texttt{GPT-4o} to refine instructions. 
These types of prompt engineering yield limited performance improvements, as shown in~\Cref{tab:runtime_instruction_impact}.
(2) We further explore advanced prompting techniques, including zero-shot CoT~\citep{kojima2022zero-shot-cot} and few-shot CoT~\citep{wei2022cot}. While zero-shot CoT enhances reasoning density and offers slight improvements (up to 4\% in the best cases), these gains remain modest. Few-shot CoT can even degrade performance, particularly for \texttt{GPT-4o-mini}, likely due to increased context length and the \textit{lost-in-the-middle} effect~\citep{liu2024lost-in-the-middle}.
(3) We also investigate prompting in a human-in-the-loop scenario to assess whether LLMs effectively leverage expert guidance in the \fillcrtext task:
\begin{packeditemize}
    \item \textbf{Distilled references:} Emulating expert guidance, we use \texttt{GPT-4o} to condense reference answers into single-sentence root cause analyses without weakness disclosure.
    \item \textbf{Enumerable directions:} From 1,922 common root causes of specification weaknesses (\eg, ``\texttt{poor failure management}''), \texttt{GPT-4o} selects the five most relevant as guidance.
\end{packeditemize}
We present examples and corresponding results in~\Cref{fig:augment-with-hints}.
The findings show that LLM like \ourmodel can benefit significantly from additional guidance, achieving up to a 78.5\% improvement in the challenging \fillcrtext task and making the augmented LLM comparable to reasoning models.
We hypothesize that expert knowledge serves as external hints, whereas reasoning models generate such hints by themselves, enabling more directed reasoning about specification weaknesses.
It is worth noting that while the distilled reference approach assumes access to preliminary high-quality expert analysis, enumerable directions remain easily accessible in production environments.
These results highlight another promising pathway for enhancing domain specialization in LLMs by incorporating expert knowledge.

\begin{table}[t]
    \centering
    \caption{\label{tab:runtime_instruction_impact}Impact of prompt refinement on \diffanalysistext. ``Inst.'' represents Instruction.}
    \resizebox{0.47\textwidth}{!}{
        \begin{tabular}{p{0.14\textwidth} cc cc}
    \toprule
          & \multicolumn{2}{c}{\textbf{LLaMA-3.1-8B}} & \multicolumn{2}{c}{\textbf{\ourmodel}}\\ \cmidrule(lr){2-3} \cmidrule(lr){4-5}
         &   \passatk{1} &  \passatk{5}  &   \passatk{1}  & \passatk{5} \\
     \midrule
        Default Inst.  & 27.4 & 59.8 & 106.4 & 148.4 \\ \hline
       Manual Inst.  & 22.0 & 49.8 & 106.3 & 146.3 \\
       GPT Inst. 1  & 18.5 & 45.2 & 106.4 & 150.2 \\
       GPT Inst. 2  & 22.0 & 48.3 & 105.3 & 143.7 \\
         \bottomrule
    \end{tabular}
    }
\end{table}

\subsection{Why Prompting Methods Fall Short}
As for the usage of LLMs, the naive method is to prompt advanced LLMs to solve target tasks in either zero-shot or few-shot manners.
However, this approach suffers from several critical limitations.
\ding{202} Prompting remains an art instead of a systematic science, making it challenging to easily craft effective prompts. 
\ding{203} Crafting prompts is not effort-free, while the effectiveness heavily relies on human expertise. This contradicts our objective of automated analysis. 
\ding{204} Prompts are typically model- and case-dependent, limiting their scalability and transferability across different scenarios and different models. Moreover, the effectiveness of even well-crafted prompts fundamentally depends on the underlying model's capabilities. As such, model limitations will bottleneck outcomes. 
Our fine-tuning methods directly enhance model abilities.
\ding{205} For more practical considerations, weakness analysis is a sensitive topic, and utilizing third-party LLM services introduces the risk of information leakage. Besides, LLM service providers often enforce strict regulations and may restrict or block security-related queries. 
This calls for the development and local deployment of specialized LLMs.

\subsection{What If Using Prompting Methods}\label{appx:exp-in-context-learning}
\noindent \textbf{Prompt baselines.} 
As we cannot enumerate all possibilities of prompts, we empirically show the performance of several representative prompt settings.
First, we ask one project member to rephrase the default instruction to test the impact of phrasing variance on LLM performance.
Second, we utilize \texttt{GPT-4o} to refine the default instruction provided in \oureval tasks by requesting more LLM-friendly variants, generating two stronger prompt baselines.
We follow default testing configurations with only the task instruction altered.
The results are shown in~\Cref{tab:runtime_instruction_impact}.
Comparing different prompt settings, we observe the sufficiency of the default instruction, which simply describes the task plainly.
We also notice that the performance of \ourmodel is not strongly dependent on the default instruction, which \ourmodel encounters in the training stage.
Besides the powerful ability obtained through domain-adaptive fine-tuning, another implicit benefit is that it eliminates the need for users to engage in extensive prompt engineering to achieve optimal performance.
This is evidenced by the smaller variance of \ourmodel's performance across multiple prompt settings.

\begin{figure}[t]
    \centering
    \includegraphics[width=\linewidth]{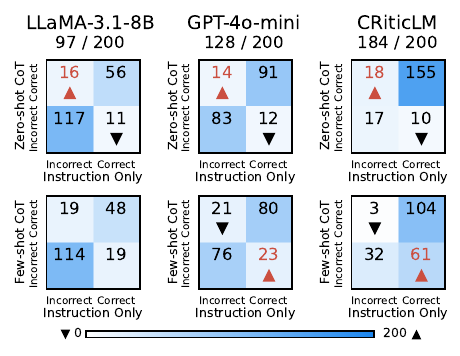}
    \captionsetup{skip=0pt}
    \caption{\label{fig:cot-impact}Impact of CoT prompting, conducted on \diffanalysistext. We use \passatk{10} to ease the sample-wise comparison and report the overall performance under diverse CoT settings for model-level comparison.}

\end{figure}

\begin{table*}[!t]
\caption{Training configurations for \ourmodel.}
\label{tab:training-config}
\centering
\newcommand{\datacolwidtha}{3.8cm}
\resizebox{0.9\textwidth}{!}{
\begin{tabular}{p{3.5cm}<{\centering}p{\datacolwidtha}<{\centering}p{\datacolwidtha}<{\centering}p{\datacolwidtha}<{\centering}}
\toprule
 & Stage 1 (DACT) & Stage 2 (TST) & Stage 3 (SCT) \\
\midrule
Corpus & \stageonedata & \stagetwodata & \stagethreedata \\
Training method & Pre-training & Supervised & Supervised \\
Learnable parameters & Full parameters & Full parameters & LoRA ($r=128, \alpha=256$) \\
Learning rate & 2e-6 & 2e-5 & 1e-4 \\
Global batch size & 256 & 128 & 64 \\
Weight decay & 0\% & 10\% & 0\% \\
Gradient clipping & 1.0 & 1.0 & 1.0 \\
Training epoch & 1 & 1 &  1 \\
Parameter precision & BF16 + TF32 & BF16 + TF32 & BF16 + TF32 \\
Warmup ratio & 10\% & 3\% & 3\% \\
Scheduler type & Cosine & Cosine & Cosine\\
Max sample length & 512 & 12,000 & 12,000\\
\bottomrule
\end{tabular}
}
\end{table*}

\noindent \textbf{CoT prompting.}
We explore the impact of recognized reasoning-enhancing techniques.
Specifically, we evaluate two representative approaches, few-shot CoT~\citep{wei2022cot} and zero-shot CoT~\citep{kojima2022zero-shot-cot}, applying them to \texttt{LLaMA-3.1-8B}, \texttt{GPT-4o-mini}, and \ourmodel.
We instantiate few-shot CoT with three randomly sampled training samples, each with augmented rationales.
As shown in~\Cref{fig:cot-impact}, the incorporation of zero-shot CoT enhances reasoning density, leading to higher pass rates across all three models.
We observe that introducing a few-shot CoT may adversely impact the performance of models.
The task instances typically span long context, \eg, the 3-shot setting additionally costs 8,287 tokens of \texttt{GPT-4o-mini}.
We hypothesize that the performance degradation stems from the \textit{lost-in-the-middle} phenomenon inherent in LLMs~\citep{liu2024lost-in-the-middle}, where long or irrelevant context will lead LLMs to behave worse.
Besides, the inference cost also increases with additional shots.

\section{Complementary Experimental Setup}\label{appx:exp-setup}

\subsection{Curating Domain-Specifc Datasets}
\label{appx:dataset-curation}

\subsubsection{\oureval}\label{subsubsec:data-stage23-eval}

\noindent \textbf{Collection.}
We first query the official database\footnote{\url{https://www.3gpp.org/ftp/Information/Databases/}} to obtain a complete list of CRs, filtering only those approved by \textit{TSGs} to ensure content reliability. 
A parallel crawler queries the CR search service\footnote{\url{http://netovate.com/cr-search/}}, retrieves FTP paths, and downloads raw CR files from the 3GPP FTP server\footnote{\url{https://www.3gpp.org/ftp}}, yielding 205,374 valid CRs.
Revisions within \textit{doc/docx} files are tracked via Office Word’s \textit{Track Changes} mode. 
Despite format evolution, 3GPP has maintained standardized CR coversheets. 
We implemented a parsing script to extract key elements, discarding change requests that failed to process. This resulted in 189,904 structured change requests.

\noindent \textbf{Annotating security relevance.}
We implement an LLM-based process to annotate the security relevance of CRs based on expert rationales \reasontext.
Using \texttt{LLaMA-3.1-70B} with the instruction in~\Cref{box:evaluate-security-relatedness}, this approach identifies 4,869 security-related CRs.
To ensure precision, we cross-reference these annotations with security-related CRs documented by~\citep{chen2022creek}, yielding 529 intersecting cases.
This is followed by a manual verification, as a small portion of CRs prove unsuitable as task instances.
For example, CRs proposing entirely new statements have empty original statements \originalstat, rendering them invalid for the \fillcrtext task.
Ultimately, we curate a set of 200\footnote{The success of compact benchmarks like HumanEval~\citep{chen2021oai-codex-and-humaneval} (164 examples) and GPQA~\citep{rein2023gpqa} (448 examples) demonstrates that small benchmarks can be effective, with faster and cheaper evaluation as an additional advantage.} high-quality security-related CRs for \oureval.

\begin{figure}[t!]
    \centering
    \includegraphics[width=1\linewidth]{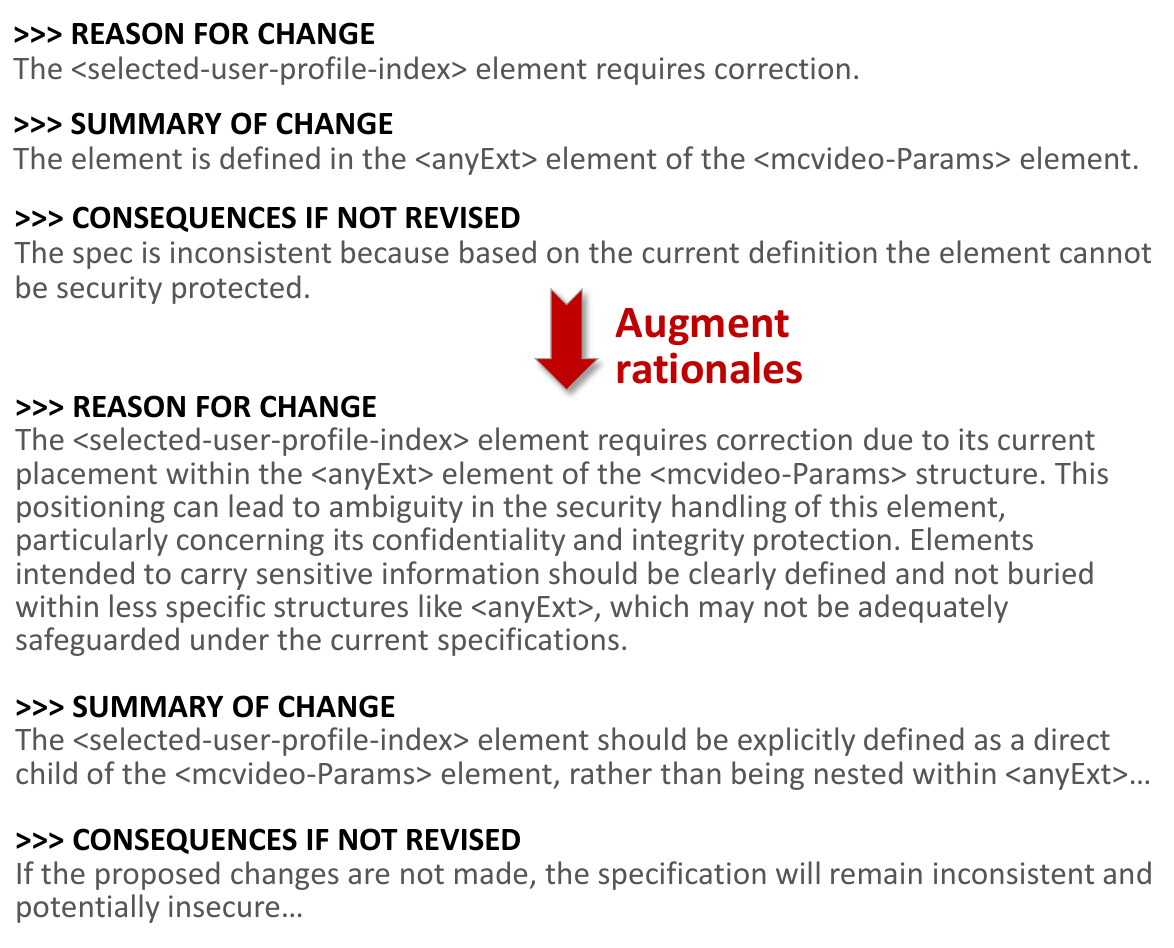}
    \vspace{-2em}
    \caption{\label{fig:rationale-augmentation-example}Example of rationale augmentation.}
    \vspace{-1em}
\end{figure}

\subsubsection{\stagetwodata and \stagethreedata}
After obtaining security-related CRs for evaluation, we have 185,035 security-unrelated CRs and 4,669 security-related CRs for the training set.
We convert all the change requests into task instances based on task formatting templates, which we introduce in~\Cref{subsec:critic-tasks}.

\noindent \textbf{Decontamination.} 
To precisely reflect the benefits of domain-adaptive fine-tuning, we try our best to minimize leakage of test cases in \oureval.
Following established practices~\citep{brown2020gpt3, achiam2023gpt-4-technical-report, radford2019gpt2}, we employ a rigorous and proactive decontamination strategy at the level of task instances.
We exclude training samples that exhibit 20-gram overlaps with any test case answers, where a gram is defined as a lowercase, whitespace delimited word.
This approach prevents both direct test case leakage and the occurrence of suspicious task instances.
Furthermore, we remove training samples associated with existing attacks discussed in~\Cref{subsec:known-vulns} using the same 20-gram matching criterion.
From this point on, the task instances of \oureval are frozen and isolated.

\noindent \textbf{Cleaning.}
Invalid task instances, as discussed in processing \oureval, are also present in the training set.
To address this, we down-sample task instances through a two-step filtering process.
First, we exclude invalid instances based on heuristic rules (\eg, extremely short queries and missing task placeholders).
Second, inspired by~\citep{gunasekar2023microsoft-phi-1, dubey2024llama-3-1-meta-llm-advance-ref-3, chen2024alpagasus-data-cleaning-ref-1}, we implement another semantic filtering to remove low-quality samples and those irrelevant to specification weaknesses.
We use \texttt{LLaMA-3.1-70B} to evaluate their educational value for specification analysis, following~\Cref{box:locate-educational-samples}.
Instances deemed to lack educational value are removed, and the remaining samples constitute our \stagetwodata dataset.
We clone security-related samples to create \stagethreedata, comprising three subsets, each aligned with a specific domain task.

\subsubsection{\stageonedata}\label{subsubsec:data-stage1}
We incorporate \textbf{3GPP standards} to enhance the LLM's comprehension of cellular networks.
Concretely, we utilize the \textit{python-docx} library to extract the main body of 2,445 specifications from the TSpec-LLM dataset~\citep{nikbakht2024tspecllmopensourcedatasetllm}. 
These specifications, spanning the 21 to 55 series and ranging from Release 8 to Release 19, cover essential aspects of cellular networks. 
We retain tables and figure captions while omitting figures due to intractability.
We also borrow a general-domain reasoning enhancement dataset, the \textbf{CoT collection dataset}~\citep{kim2023cot-collection-dataset-rationalization-ref-3}.
To mitigate catastrophic forgetting~\citep{scialom2022fine-tuned-lm-continual-learning-replay-ref-1}, we include the \textbf{Wikipedia dataset}~\citep{0-wikidump-llm-dataset} and the \textbf{ArXiv split} from the RedPajama dataset~\citep{weber2024redpajama-pretraining-dataset}.
We filter these general-domain datasets using keyword-based heuristics to identify documents specifically relevant to cellular networks and security, ensuring focused domain adaptation.

\subsection{Training Configurations}
\label{appx:training-configs}

We list main training configurations in~\Cref{tab:training-config}.
We choose a high rank $r=256$ for the LoRA adapter and set $\alpha=2r$ as recommended by~\citet{ biderman2024lora-learn-less-and-forget-less}. 
We apply LoRA adapters to all linear layers in the model.
All three training stages employ AdamW optimizer~\citep{loshchilov2017adamw} with $\beta_1$ as 0.9 and $\beta_2$ as 0.999.
The settings of learning rates are $2 \times 10^{-6}$ for DAPT, $2 \times 10^{-5}$ for TST, and $1 \times 10^{-4}$ for SCT.
For batch size, we employ 256 for DAPT, 128 for TST, and 64 for SCT.
All the training stages consume one epoch with the analogous learning scheduler: the learning rate is linearly warmed up for several training steps, and then cosine-decreases to $1/20$ of the peak learning rate.
Gradient accumulation is adopted to achieve large batch sizes with constrained GPU memory.

\noindent \textbf{Compute infrastructure.}
All experiments were conducted on a server running Ubuntu 20.04.5 LTS operating system.
The machine is equipped with an Intel Xeon Platinum 8468V processor (96 cores, 192 threads), 2 TB of system memory, and 8 NVIDIA H800 GPUs with 80 GB of VRAM each.

\noindent \textbf{Software.}
Our project is implemented based on Python 3.12, CUDA 11.8, PyTorch 2.4.0, and HuggingFace's transformer library. 
To accelerate training, we achieve data parallel through DeepSpeed~\citep{rajbhandari2020zero-deepspeed}: we adopt ZeRO stage-2 with a world size of 4 for 8B models and stage-3 with a world size of 8 for 70B models.
We use Flash-Attention 2~\citep{dao2024flashattention-2} to improve throughput and use gradient checkpointing to reduce memory requirements. 
For evaluation, we deploy inference endpoints using vLLM~\citep{kwon2023vllm}. 
The entire project consumes around 12,800 lines of code, decomposed in~\Cref{tab:code-lines}.

\begin{table}[t]
\centering

\caption{\label{tab:evaluated-models}Models evaluated in this work.}
\resizebox{\linewidth}{!}{\begin{tabular}{lccc}
\toprule
Model & Model size & Open-source & Context window \\
\midrule
GLM-4-9B            & 9B        & Y     & 128K      \\
GLM-4.5            & MoE & Y &  128K    \\
Mistral-7B-v0.3     & 7B        & Y     & 128K      \\
InternLM-2.5-7B     & 7B        & Y     & 1M        \\
Qwen-2.5-7B         & 7B        & Y     & 128K      \\
Qwen-2.5-14B         & 14B        & Y     & 128K      \\
Qwen-2.5-32B         & 32B        & Y     & 128K      \\
Qwen-2.5-72B         & 72B        & Y     & 128K      \\
Qwen3-8B            &  8B & Y &  128K    \\
Qwen3-14B            & 14B & Y &   128K   \\
Qwen3-32B            & 32B & Y &  128K    \\
Qwen3-30B-A3B       & MoE & Y &   128K   \\
Qwen3-235B-A22B      &  MoE & Y &   128K   \\
DeepSeek-V3     & MoE  & Y &    128K  \\
DeepSeek-R1       & MoE       & Y     & 128K      \\
LLaMA-3.1-70B       & 70B       & Y     & 128K      \\
LLaMA-3.1-8B        & 8B        & Y     & 128K      \\
\midrule
Doubao-seed-1-6-flash            & Unknown & N &   256K   \\
Doubao-seed-1-6-lite            & Unknown & N &   256K   \\
Doubao-seed-1-6            & Unknown & N &  256K    \\
Doubao-seed-1-6-thinking            & Unknown & N &    256K  \\
Claude-3.5-Sonnet   & Unknown   & N     & 200K      \\
Gemini-2.0-flash-thinking         & Unknown   & N     & 1M      \\
Gemini-2.5-flash            & Unknown & N & 1M     \\
Gemini-2.5-pro            & Unknown & N &   1M   \\
GPT-3.5-turbo       & Unknown   & N     & 16,385    \\
GPT-4o-mini         & Unknown   & N     & 128K      \\
GPT-4o              & Unknown   & N     & 128K      \\
GPT-o3-mini       & Unknown   & N     & 200K    \\
GPT-5-mini           & Unknown & N &   400k   \\
GPT-5           & Unknown & N &   400k   \\
\bottomrule
\end{tabular}}

\end{table}

\begin{table}[t]
\centering
\caption{\label{tab:code-lines}Module-level lines of code counted using \textit{cloc}.}
\setlength\tabcolsep{8pt}
\small
\resizebox{0.85\linewidth}{!}
{\begin{tabular}{p{0.6\linewidth}C{0.1\linewidth}}
    \toprule
    \textbf{Component}              & \textbf{LoC} \\
    \midrule
    Data collection and processing  & 4,946  \\
    Training                        & 3,419  \\
    \oureval                       & 2,203  \\
    User study                      & 2,241  \\
    \midrule
    \textbf{Total}                  & 12,809 \\
    \bottomrule
\end{tabular}}

\end{table}

\section{Additional Experiment Results}\label{appx:additional-exps}

\subsection{Correlation between \oureval Tasks}\label{appx:task-correlation}
We explore the correlation between the \oureval tasks by evaluating whether the knowledge acquired by training on the source task can transfer to the target task.
We train base models with \stagethreedata of the source task and evaluate the trained model in the target task.
The results are shown in~\Cref{tab:task-correlation}.
We notice that the knowledge is clearly transferable between tasks, which substantiates the efficacy of our multi-task learning design in the TST stage.

\begin{table}[H]
\centering
\caption{Cross-task performance: models are trained on source tasks with \stagethreedata and evaluated on target tasks using \passatk{5}.}
\label{tab:task-correlation}
    \resizebox{0.95\columnwidth}{!}{
    \begin{tabular}{c|l|ccc}
        \toprule
        \multicolumn{2}{c|}{\multirow{2}{*}{}} & \multicolumn{3}{c}{\textbf{Source}} \\
        \cmidrule{3-5}
        \multicolumn{2}{c|}{} & \textbf{\outlinerevtitle} & \textbf{\diffanalysistitle} & \textbf{\fillcrtitle} \\
        \midrule
        \multirow{3}{*}{\rotatebox[origin=c]{90}{\textbf{Target}}} 
        & \outlinerevtitle   & \cellcolor{gray!15}177.9 & 183.6 & 169.8 \\
        & \diffanalysistitle & 77.8  & \cellcolor{gray!15}124.2 & 74.4 \\
        & \fillcrtitle       & 31.9  & 27.2  & \cellcolor{gray!15}33.5 \\
        \bottomrule
    \end{tabular}
    }

\end{table}

\begin{figure*}[!t]
    \centering
    \begin{subfigure}[b]{0.32\textwidth}
    \includegraphics[width=\textwidth]{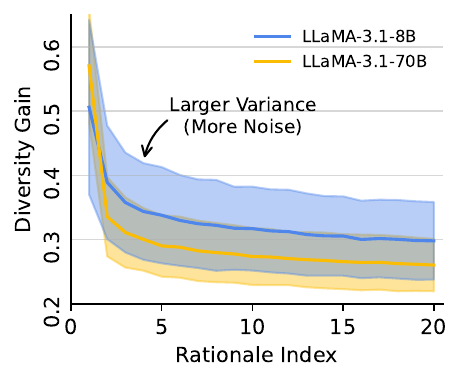}
    \label{fig:diversity-gain-8b-70b}
    \end{subfigure}
    \hfill
    \begin{subfigure}[b]{0.32\textwidth}
        \includegraphics[width=\textwidth]{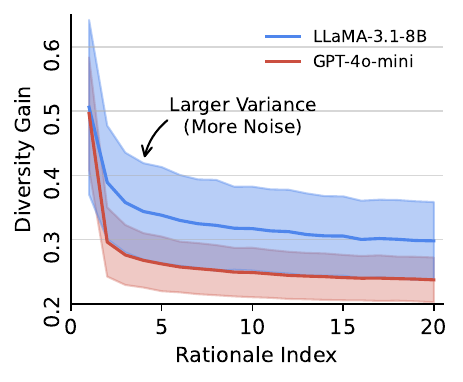}
        \label{fig:diversity-gain-8b-4o-mini}
    \end{subfigure}
    \hfill
    \begin{subfigure}[b]{0.32\textwidth}
        \raisebox{0.0em}{\includegraphics[width=\textwidth]{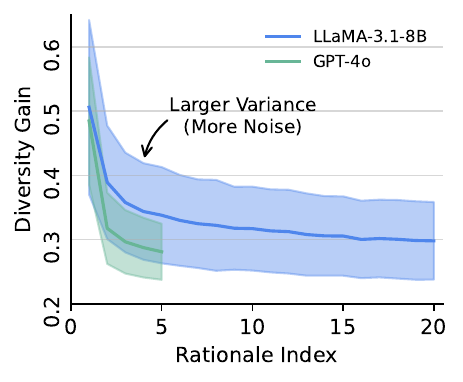}}
        \label{fig:diversity-gain-8b-4o}
    \end{subfigure}

    \captionsetup{skip=-3pt}
    \caption{Patterns of diversity gain when augmenting rationales using different models: \texttt{LLaMA-3.1-8B}, \texttt{GPT-4o-mini}, \texttt{GPT-4o}, and \texttt{LLaMA-3.1-70B}. The rationales correspond to the \diffanalysistext task. We use Euclidean distance to measure the similarity between rationales while the sample-level diversity gain is measured as the minimal distance of the $i$-th rationale against the set of the previous $(i-1)$ rationales plus the original answer. We plot the variance across different task instances.}
    \label{fig:diversity-gain-combined}
    
\end{figure*}

\subsection{Tracking Model Behaviors}\label{appx:track-behavior}

We study why \ourmodel can excel in cellular specification refinement and, consequently, on \oureval.
We analyze the model's behavior by collecting next-token predictions during processing the \diffanalysistext task of \oureval.
Formally, we obtain a set of softmax-normalized next-token prediction probabilities denoted as ${p_i^j\in\mathbb{R}^{|V|}, i\in[1,S_j], j\in[1,N]}$, where $N$ denotes the number of test cases in \oureval, $S_j$ represents the sequence length of predicted tokens for the $j$-th test case under greedy search, and $|V|$ is the vocabulary size.
To mitigate varying completion lengths, we perform hierarchical aggregation: computing the mean prediction distribution within each sample ($\frac{1}{S_j}\sum_{i=1}^{S_j} p_i^j$), and then averaging across all $N$ samples.
This processing condenses the LLM's behavior on \oureval into a single probability distribution $P_{\text{LLM}}\in\mathbb{R}^{|V|}$, where each dimension represents the model's averaged behavior for a vocabulary token.
Due to the huge vocabulary size, \eg, $|V|=131,072$ for \texttt{LLaMA-3.1} models, we focus on tokens with probabilities higher than $\frac{1}{|V|}$, which represent frequently used vocabulary in LLM outputs.
We conduct a comparative analysis between $P_{\ourmodel}$ and $P_{\text{LLaMA-3.1-8B}}$, with key observations presented in~\Cref{tab:token-behavior-full}.
Significantly, \ourmodel yields higher probabilities for security-related tokens, and notably transitions from employing generic descriptions (\eg, ``\_errors'' and ``\_risks'') to more specific terminologies (\eg, ``\_interception'' and ``\_confidentiality'').
As the $P_{\text{LLM}}$ is normalized, an increase in certain token probabilities inevitably results in the reduction of others.
This transition aligns with our objective of developing a more domain-specialized model.

\subsection{Exploring Diversity Gain of Rationales}\label{appx:explain-rationale}

Building upon our scalability analysis rationales in~\Cref{subsubsec:scalability}, we study why training with more rationale-augmented answers can benefit \ourmodel and why this improvement finally plateaus.
We analyze the semantic differences between rationales by measuring their distances in the embedding space.
We employ a feature extractor (\ie, OpenAI's text-embedding-3-large) to project each rationale into the embedding space.
We then observe the diversity gain brought by progressively adding new rationales.
The diversity gain corresponding to the $i$-th rationale of the $j$-th task instance is defined as the minimum distance between the new rationale and the union of existing rationales and the original answer, formalized as:
$\min_{k < i} {|r_{i,j} - r_{k,j}|_2 \cup |r_{i,j} - a_j|_2}$
where $r_{i,j}$ denotes the $i$-th rationale for the $j$-th task instance, $a_j$ is the original answer, and $|\cdot|_2$ represents the Euclidean distance in the embedding space.
Our analysis in~\Cref{fig:diversity-gain-combined} reveals that the marginal diversity gain per new rationale diminishes as the number of rationales increases.
This trend correlates with the observed improvement of \ourmodel as we increase the rationale number.
We stipulate that this convergence occurs because additional rationales fail to introduce new insights, and the remaining diversity gains stem from the altered wordings.
Notably, when using \texttt{LLaMA-3.1-8B} as the rationale generator, we observe diversity gains with both a higher mean and significantly greater variance compared to other models.
This observation partially explains the inferior performance of \texttt{LLaMA-3.1-8B} as the rationale generator and provides insights for choosing rationale generators.

\begin{table}[t]
    \centering
    \caption{Impact of training stages, with rigorous decontamination to minimize memorization.}
    \label{tab:stage_design}
    \renewcommand{\arraystretch}{1.06}
    \resizebox{0.47\textwidth}{!}{%
    \begin{tabular}{p{0.16\textwidth} cc cc}
        \toprule
          & \multicolumn{2}{c}{\textbf{\diffanalysistitle}} & \multicolumn{2}{c}{\textbf{\fillcrtitle}}\\ \cmidrule(lr){2-3} \cmidrule(lr){4-5}
         &   \passatk{1} &  \passatk{5}  &   \passatk{1}  & \passatk{5} \\
     \midrule
       LLaMA-3.1-8B  & 27.4 & 59.8 & 6.1 & 18.1 \\ \hline
       \multicolumn{1}{r}{+ SCT}  & 73.3 & 124.2 & 12.7 & 33.5 \\
       \multicolumn{1}{r}{+ TST + SCT}  & 95.8  & 145.3  &  25.3 & 49.9 \\
       \multicolumn{1}{r}{+ DACT + TST + SCT}  & 106.4  & 148.4  & 27.2 & 57.8 \\
         \bottomrule
    \end{tabular}
    }
\end{table}

\subsection{Ablating Training Stages}\label{appx:ablation-study}
To assess the contributions of each training stage, we conduct an ablation study.
As shown in~\Cref{tab:stage_design}, each stage positively contributes to \ourmodel's performance on \oureval tasks.
The SCT stage, closely aligned with security-centric analysis, yields the most significant improvement.
Training with security-irrelevant samples in TST also enhances performance by improving generalization in addressing specification weaknesses.
The DACT stage, designed to complement LLMs with domain knowledge, provides modest gains, likely because base LLMs (\eg, \texttt{LLaMA-3.1-8B}) already possess relevant knowledge~\citep{zhou2024lima}.

\subsection{Scaling Along the Data Dimension}\label{appx:data-dim-scalability}
We examine how model performance evolves with increasing training data volume.
We focus on the \diffanalysistext task, as it is the most distinguishing task in \oureval.
Training data scales approximately logarithmically, with each data point representing a full training run using default hyperparameters.
As illustrated in~\Cref{fig:Scalability-data}, results reveal two key trends: 1) performance consistently improves with more data, and 2) performance gains exhibit diminishing returns, aligning with observations of established scaling laws~\citep{kaplan2020scaling, hoffmann2022Chinchilla-scaling-law, dubey2024llama-3-1-meta-llm-advance-ref-3}.

\subsection{Extensibility to Advanced Base Models}
\label{appx:sft-other-models}

\begin{figure}[!t]
    \centering

    \includegraphics[width=1\linewidth]{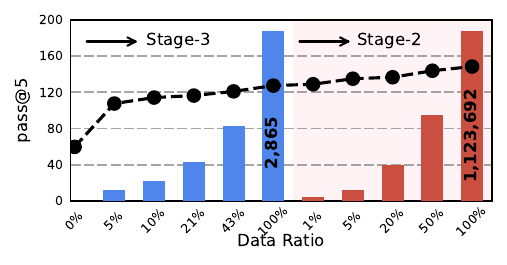}
    \vspace{-2em}
    \caption{\label{fig:Scalability-data}Data-dimension scalability.}
    \vspace{1em}
    \includegraphics[width=\linewidth]{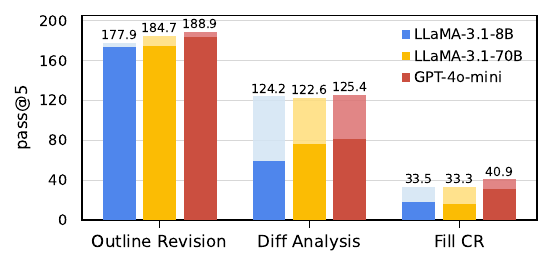}
    \vspace{-2.5em}
    \caption{\label{fig:exp-param-scaling}Extensibility to advanced base models.}

    \vspace{-1em}
\end{figure}

In our experiments, we primarily focus on the LLM with a ``small'' parameter count, LLaMA-3.1-8B.
We extend the domain-adaptive fine-tuning to closed-sourced \texttt{GPT-4o-mini}\footnote{Fine-tuning GPT models is officially accessible via \url{https://platform.openai.com/finetune}. Considering cost affordability, we select GPT-4o-mini as a representative example of closed-source models.} and the \texttt{LLaMA-3.1-70B} with a larger parameter count, using the \stagethreedata.
As shown in~\Cref{fig:exp-param-scaling}, domain-adaptive fine-tuning yields performance improvements across all models.
Interestingly, \texttt{LLaMA-3.1-8B} and \texttt{LLaMA-3.1-70B} converge to similar levels, indicating that domain data quality, rather than model size, is the primary bottleneck in certain cases.
Meanwhile, \texttt{GPT-4o-mini} consistently maintains superior performance, particularly on the \fillcrtext task.
This suggests that applying our training methodology to advanced foundation models could further specialize LLMs.

\subsection{Examining Existing Attacks}\label{appx:known-vulns}

\noindent \textbf{Considered threat models.}
The settings of the threat model strictly follow their corresponding original papers.
In gross, we consider both passive and active attacker models. 
\begin{packeditemize}
\item \textbf{Passive adversary:} This attacker can eavesdrop on over-the-air radio broadcast channels, such that they can analyze and deduce information from intercepted messages.
\item \textbf{Active adversary:} This attacker can establish and operate a rogue base station to inject malicious traffic directed at UEs. While they are assumed to have full knowledge of the protocol specifications, they lack access to cryptographic keys, except for public keys.
\end{packeditemize}
For certain scenarios, we suppose the attacker also knows some identity information of the victim UE, like the C-RNTI. Alternatively, the attacker might have a hypothesis about the victim’s identity information and seek to verify it.

\section{Structural Analysis of Datasets}\label{appx:structural-analysis}
We demonstrate the representativeness of \oureval for evaluating cellular specification refinement.
We also provide statistics about the training dataset size.

\begin{figure*}[t]
    \centering
    \captionsetup{skip=-3pt}
    
    \begin{minipage}{0.48\linewidth}
        \centering
        \includegraphics[width=\linewidth]{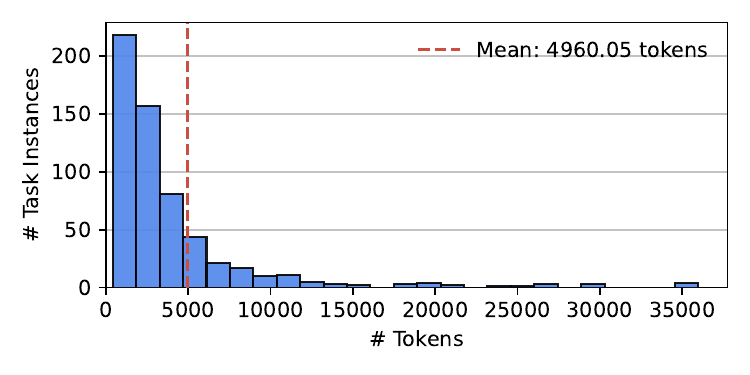}
        \caption{Distribution of token counts of test cases in \oureval, with three tasks combined.}
    
        \label{fig:token-number-dist}
    \end{minipage}
    \hfill
    \begin{minipage}{0.48\linewidth}
        \centering
        \includegraphics[width=\linewidth]{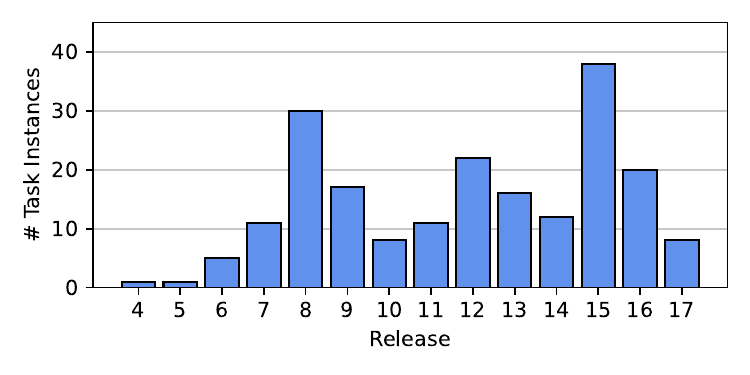}
        \caption{Release distribution of CRs in \oureval.}
    
        \label{fig:release-dist}
    \end{minipage}
    
\end{figure*}

\noindent \textbf{Sample length of \oureval.}
We analyze the sample length distribution of test cases in \oureval at the token level, as demonstrated in~\Cref{fig:token-number-dist}. 
The test cases of \oureval typically contain thousands of or even tens of thousands of tokens, presenting rigorous challenges that specifically test models' long-context capabilities.
Moreover, LLMs suffer the problem of \textit{lost-in-the-middle}, meaning that the models claiming long context cannot effectively leverage the information given in their context~\citep{liu2024lost-in-the-middle}.
Models without sufficiently effective long context (less than the claimed maximum token number) cannot tackle the test cases in a single inference.

\begin{figure}[!t]
\centering
\hspace{-1cm} 
\includegraphics[width=0.9\linewidth]{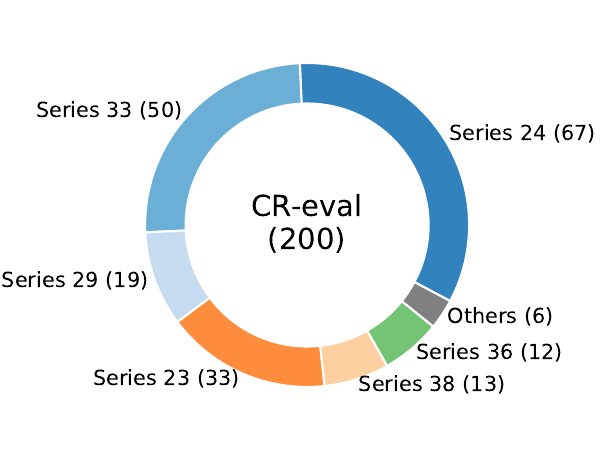}
\vspace{-1em}
\label{fig:spec-dist}
\caption{ Distribution of CRs in \oureval across different specification series.}
\vspace{-1em}
\end{figure}

\begin{table*}[!t]
\centering
\hspace{-1cm} 
    \centering
    \begin{tabular}{lrrr}
    \toprule
         & \# Samples & \# Tokens & \# Response tokens \\
    \midrule
    \oureval (\outlinerevtext)  & 200 & 898,521 & 15,642 \\
    \oureval (\diffanalysistext) & 200 & 1,194,454 & 73,239 \\
    \oureval (\fillcrtext)  & 200 &  883,054 & 73,239 \\ \hline
    \stageonedata (shared)   & 3,729,713 & 1,433,683,482 & 1,433,683,482 \\
    \stagetwodata (shared)  & 1,123,692 & 4,892,550,046 & 661,498,363 \\
    \stagethreedata (\outlinerevtext) & 13,860 & 44,085,131 & 3,087,016 \\
    \stagethreedata (\diffanalysistext)  & 14,325 & 56,114,617 & 5,427,157 \\
    \stagethreedata (\fillcrtext)   & 15,370 & 53,544,878 & 6,305,663 \\
    \bottomrule
    \end{tabular}
    \label{tab:data-structure}

\caption{Token statistics of the datasets at both the sample and token level, based on the tokenizer of the \texttt{LLaMA-3.1} herds~\citep{dubey2024llama-3-1-meta-llm-advance-ref-3}.}
\end{table*}

\begin{table}[t]
    \centering
    \vspace{-1em}
    \caption{Full list of \ourmodel's token prediction behavior on \oureval. The ratios are relative to the base model (LLaMA-3.1-8B). Note: '\_' denotes the blank character in tokens.}
    \label{tab:token-behavior-full}
    \Large
    \renewcommand{\arraystretch}{0.97}
    \resizebox{0.46\textwidth}{!}{
        \begin{tabular}{cccc}
    \toprule
    Token & Ratio & Token & Ratio \\
    \midrule
       \_safeguard & $138.70 \times$ & \_degrade & $79.21 \times$ \\
       \_improper & $58.78 \times$ & Failure & $61.47 \times$ \\
       \_mistakenly & $39.49 \times$ & \_challenges & $47.90 \times$ \\
       \_interception & $32.64 \times$ & \_interruptions & $24.07 \times$ \\
       \_inadvertently & $21.85 \times$ & \_reuse & $16.60 \times$ \\
       \_misuse & $13.98 \times$ & \_operational & $16.38 \times$ \\
       \_susceptible & $12.34 \times$ & \_cryptographic & $10.98 \times$ \\
       \_legal & $10.49 \times$ & \_degraded & $10.07 \times$ \\
       \_disrupt & $6.90 \times$ & \_misunderstanding & $6.5 \times$ \\
       \_unintended & $6.77 \times$ & \_privacy & $5.43 \times$ \\
       \_fail & $4.66 \times$ & \_protecting & $4.66 \times$ \\
       \_unprotected & $4.31 \times$ & \_risk & $3.82 \times$ \\
       \_ambiguity & $3.70 \times$ & \_invalid & $3.73 \times$ \\
       \_spoof & $3.41 \times$ & \_trust & $3.62 \times$ \\
       \_reliability & $3.29 \times$ & \_disruptions & $3.10 \times$ \\
       \_incorrectly & $2.88 \times$ & \_confidentiality & $2.96 \times$ \\
       \_legitimate & $2.84 \times$ & \_protected & $2.86 \times$ \\
       \_unauthorized & $2.62 \times$ & \_compromise & $2.83 \times$ \\
       \_integrity & $2.62 \times$ & \_dereg & $2.56 \times$ \\
       \_leaks & $2.51 \times$ & \_ambigu & $2.36 \times$ \\
       \_breaches & $2.35 \times$ & \_compliance & $2.34 \times$ \\
       \_intercepted & $2.27 \times$ & \_manipulation & $2.32 \times$ \\
       \_authenticity & $2.14 \times$ & \_inconsistency & $2.23 \times$ \\
       \_disruption & $1.97 \times$ & \_threats & $2.11 \times$ \\
       \_rejection & $1.85 \times$ & \_degradation & $1.95 \times$ \\
       \_securely & $1.77 \times$ & \_vulnerability & $1.77 \times$ \\
       \_lack & $1.67 \times$ & \_ambiguous & $1.70 \times$ \\
       \_safety & $1.63 \times$ & \_authenticated & $1.66 \times$ \\
       \_failures & $1.62 \times$ & \_robust & $1.61 \times$ \\
       \_attacks & $1.56 \times$ & \_interoper & $1.54 \times$ \\
       \_compromising & $1.48 \times$ & \_inability & $1.52 \times$ \\
       \_authorized & $1.46 \times$ & \_malicious & $1.47 \times$ \\
       \_failed & $1.40 \times$ & \_intended & $1.46 \times$ \\
       \_unable & $1.32 \times$ & \_Privacy & $1.34 \times$ \\
       \_incorrect & $1.27 \times$ & \_consistency & $1.23 \times$ \\
       \_availability & $1.21 \times$ & \_compromised & $1.22 \times$ \\
       \_authenticate & $1.15 \times$ & \_confusion & $1.14 \times$ \\
       \_secure & $1.11 \times$ & \_authentication & $1.14 \times$ \\
       \_security & $1.07 \times$ & \_consistent & $1.05 \times$ \\
       \_predictable & $0.96 \times$ & \_unexpected & $0.97 \times$ \\
       \_attacker & $0.91 \times$ & \_certificate & $0.91 \times$ \\
       \_inconsistencies & $0.86 \times$ & \_disabled & $0.88 \times$ \\
       \_authorization & $0.81 \times$ & \_error & $0.81 \times$ \\
       \_vulnerabilities & $0.81 \times$ & \_inconsistent & $0.80 \times$ \\
       \_vulnerable & $0.72 \times$ & \_errors & $0.71 \times$ \\
       \_risks & $0.64 \times$ & \_comply & $0.54 \times$ \\
       \_unclear & $0.53 \times$ & \_reliable & $0.52 \times$ \\
       \_barred & $0.45 \times$ & \_compatibility & $0.43 \times$ \\
       \_flexibility & $0.42 \times$ & \_negative & $0.41 \times$ \\
       \_damage & $0.40 \times$ & \_difficulties & $0.20 \times$ \\
    \bottomrule
    \end{tabular}
    }
\end{table}

\noindent \textbf{Release coverage of \oureval.}
We provide the statistics of the target releases of CRs in \oureval, which is illustrated in~\Cref{fig:release-dist}.
\oureval demonstrates extensive release coverage and excellent diversity, with its involved CRs spanning from Release 5 to Release 17.
Covering CRs of old releases in \oureval is necessary as historical security vulnerabilities can provide insights for refining contemporary cellular specifications.
Intriguingly, we observed more security-related CRs during Release 8 and Release 15, which correspond to the introductions of LTE and 5G, respectively~\citep{chen2022creek}.
This underscores the importance of automated cellular specification refinement methods, particularly during major technological transitions.
To ensure rigorous identification of security-relevant CRs, we cross-referenced our annotations with those from \citet{chen2022creek}.
While this early decision enhanced the reliability of our security relevance annotations, it also brought about an unexpected consequence, specifically, \oureval currently excludes test cases from Release 18 onward. 
Future work will address this by developing an extended version of \oureval incorporating more recent CRs.

\noindent \textbf{Specification coverage of \oureval.}
\oureval encompasses 200 CRs distributed across 74 distinct specifications, demonstrating its extensive scope.
We provide a coarse-grained summary of the specification distribution according to the belonging to standard series in~\Cref{fig:spec-dist}.
Unlike previous works that primarily focus on a limited set of specifications such as NAS and RRC, \oureval provides comprehensive coverage across the 3GPP ecosystem.
Yet the broad coverage makes it impractical for us to establish a human-level baseline on \oureval.

\noindent \textbf{Token number of all datasets.}
We present the dataset decomposition in~\Cref{tab:data-structure}.
Note that we count the token number of data after rationale augmentation.
As the auto-regressive LLMs are trained through the next token prediction task~\citep{radford2018gpt1}, the dataset size at the token level can more precisely show how much the LLMs can learn from the training.
For the continual pre-training paradigm, \eg, DACT in our framework, all tokens are learnable while only the response tokens can be learned for the supervised fine-tuning paradigm.
Scaling law~\citep{kaplan2020scaling, hoffmann2022Chinchilla-scaling-law, muennighoff2023scaling-data-constrained-llm-training} demonstrates that LLMs can consistently gain benefits through continual training investment.
An implicit side of the scaling law is what the training dataset teaches the model.
That's the rationale behind our finding that a limited number of security-related domain data \stagethreedata contributes significantly to the performance improvement on \oureval.
This underscores the crucial importance of developing high-quality domain datasets closely relevant to our target task, cellular specification refinement.

\noindent \textbf{Difficulty of \oureval.}
The difficulty of test cases can be naturally measured through their \textit{global solving rate}, defined as the proportion of model trials capable of solving them. 
We reuse the model predictions in~\Cref{subsec:evaluation-results} and aggregate the solving rates across all model trials (10 trials per model).
We provide the statistics in~\Cref{fig:solving-rates-of-benchmark}.
The final \oureval aligns with our design principle of progressive challenge levels. 
At a macro level, the three tasks exhibit distinct difficulty tiers, as evidenced by their mean solving rates.
For example, the \fillcrtext task presents the highest challenge by providing the model with minimal information while demanding an in-depth understanding of potential weaknesses within the specification clauses.
At a micro level, each task comprises test cases of varying difficulty, as substantiated by our demonstrations in~\Cref{subsec:evaluation-results}.
While all test cases passed our manual verification process, ensuring that they provide sufficient information for task completion, they incorporate different implicit confounding factors, \eg, the provided context specification statements and the expected response quality.
Among the three tasks, both the \diffanalysistext and \fillcrtext are challenging enough to differentiate LLMs' domain-specific abilities, despite rapid advances in LLM development.

\section{Human Study for LLM-as-a-Judge}\label{appx:human-study-details}
We conducted the human study with eight PhD students specializing in network security. 
Their research experience in the field ensures the quality of our evaluation.
All participants volunteered and were willing to contribute their annotations to the community.
The human study concerning the reliability of LLM-as-a-Judge consists of two rounds, whose system snapshots are presented in~\Cref{fig:human-study-system-snapshot-1} and~\Cref{fig:human-study-system-snapshot-2}:
1) Alignment test: 
Participants were presented with 25 samples, each consisting of an LLM response and the corresponding reference answer. 
Participants are tasked to \textit{accept} or \textit{reject} the LLM responses based on their alignment with the reference answers. 
This round evaluates Human-as-a-Judge under the fair setting with LLM-as-a-Judge, aiming to test their alignment.
2) Judgment approval test: 
Participants are additionally presented with the LLM-as-a-Judge's judgment and its posterior explanation. 
Participants are asked to \textit{approve} or \textit{disapprove} the LLM-as-a-Judge's judgment. 
This round aimed to calibrate the rigor degree of humans and show the acceptableness of LLM-as-a-Judge's decisions from the perspective of human annotators.

\definecolor{cblue}{HTML}{4F86EC}
\definecolor{cred}{HTML}{CB4F40}
\definecolor{light_blue}{HTML}{D3E4F4}
\definecolor{light_yellow}{HTML}{FFDE7A}
\definecolor{light_red}{HTML}{EF5A6F}
\definecolor{margin_blue}{HTML}{7EACB5}

\newcommand{\llmacc}{Accept}
\newcommand{\llmrej}{Reject}
\newcommand{\rej}{\multicolumn{2}{|c}{Reject}}
\newcommand{\acc}{\multicolumn{2}{|c}{Accept}}
\newcommand{\ator}{\multicolumn{2}{|c}{A$\to$ R}}
\newcommand{\rtoa}{\multicolumn{2}{|c}{R$\to$ A}}

\renewcommand{\arraystretch}{1.2}

\begin{table*}
    \centering
    \caption{The raw data of human study. Each column corresponds to one human annotator. Each row corresponds to one sample and we cluster the samples based on LLM-as-a-Judge's decisions for readability.} \label{tab:human-study-raw-data}
    \small
    \resizebox{0.95\textwidth}{!}{
    \begin{tabular}{C{1cm}|c|cc|cc|cc|cc|cc|cc|cc|cc}
\toprule
Idx & LLM-as-a-Judge & \multicolumn{2}{c}{1} & \multicolumn{2}{|c}{2} & \multicolumn{2}{|c}{3} & \multicolumn{2}{|c}{4} & \multicolumn{2}{|c}{5} & \multicolumn{2}{|c}{6} & \multicolumn{2}{|c}{7} & \multicolumn{2}{|c}{8}\\ 
\midrule
 1& \llmacc & \acc   & \rtoa & \rej  & \rtoa & \acc  & \acc  & \rtoa & \rtoa \\ \hline
 4& \llmacc & \acc   & \acc  & \acc  & \acc  & \rtoa & \acc  & \acc  & \rtoa \\ \hline
 5& \llmacc & \acc   & \acc  & \acc  & \acc  & \acc  & \acc  & \acc  & \acc  \\ \hline
 7& \llmacc & \acc   & \acc  & \acc  & \acc  & \acc  & \acc  & \acc  & \acc  \\ \hline
 9& \llmacc & \acc   & \acc  & \acc  & \acc  & \acc  & \acc  & \acc  & \acc  \\ \hline
11& \llmacc & \acc   & \rtoa & \rej  & \acc  & \acc  & \acc  & \acc  & \acc  \\ \hline
15& \llmacc & \acc   & \acc  & \acc  & \acc  & \acc  & \acc  & \acc  & \rtoa \\ \hline
16& \llmacc & \acc   & \acc  & \acc  & \acc  & \acc  & \acc  & \acc  & \acc  \\ \hline
18& \llmacc & \acc   & \acc  & \acc  & \acc  & \acc  & \rtoa & \acc  & \acc  \\ \hline
22& \llmacc & \acc   & \acc  & \acc  & \acc  & \acc  & \acc  & \acc  & \acc  \\ \hline
24& \llmacc & \acc   & \acc  & \acc  & \acc  & \rtoa & \acc  & \acc  & \rej  \\ \hline
25& \llmacc & \acc   & \acc  & \acc  & \acc  & \acc  & \acc  & \acc  & \acc  \\ \hline
 2 & \llmrej & \rej   & \rej  & \rej  & \rej  & \rej  & \rej  & \rej  & \rej  \\ \hline
 3 & \llmrej & \rej   & \rej  & \rej  & \rej  & \rej  & \rej  & \rej  & \ator \\ \hline
 6 & \llmrej & \rej   & \rej  & \rej  & \rej  & \rej  & \rej  & \rej  & \rej  \\ \hline
 8 & \llmrej & \rej   & \rej  & \rej  & \rej  & \rej  & \rej  & \rej  & \rej  \\ \hline
10 & \llmrej & \ator  & \rej  & \rej  & \rej  & \ator & \ator & \ator & \rej  \\ \hline
12 & \llmrej & \ator  & \rej  & \rej  & \rej  & \rej  & \rej  & \rej  & \ator \\ \hline
13 & \llmrej & \ator  & \rej  & \rej  & \rej  & \rej  & \rej  & \rej  & \rej  \\ \hline
14 & \llmrej & \rej   & \acc  & \acc  & \rej  & \rej  & \rej  & \rej  & \rej  \\ \hline
17 & \llmrej & \rej   & \acc  & \rej  & \rej  & \rej  & \ator & \rej  & \ator \\ \hline
19 & \llmrej & \rej   & \rej  & \acc  & \rej  & \ator & \ator & \ator & \ator \\ \hline
20 & \llmrej & \rej   & \rej  & \rej  & \rej  & \rej  & \rej  & \rej  & \rej  \\ \hline
21 & \llmrej & \rej   & \rej  & \rej  & \rej  & \rej  & \rej  & \rej  & \ator \\ \hline
23 & \llmrej & \ator  & \acc  & \ator & \rej  & \ator & \ator & \rej  & \rej  \\ \bottomrule
    \end{tabular}
}
\end{table*}

\begin{figure}
    \centering
    \captionsetup{skip=1pt}

    \begin{tikzpicture}
        \begin{axis}[
            width=\linewidth,
            xmajorgrids,
            grid style={dashed, line width=0.6pt},
            axis line style={draw=none},
            xtick style={draw=none},
            ytick style={draw=none},
            bar width=0.26cm,
            xmin=0,
            xmax=162,
            ymin=0,
            ymax=8,
            enlarge y limits=0.05,
            xlabel={Time consumed (min)},
            ylabel={Participant ID},
            ylabel style={xshift=6pt,yshift=-10px},
            ytick={0,1,2,3,4,5,6,7,8},
            yticklabels={LLM,1,2,3,4,5,6,7,8},
            point meta=explicit symbolic,
            nodes near coords,
            every node near coord/.append style={
                font=\scriptsize,
                anchor=west,
                text=black
            }
        ]
            \addplot [
                xbar, fill=cred, mark=none, draw=none, bar shift=0,
                node near coords style={font=\scriptsize\bfseries}
            ] coordinates {(0.6,0) [36 s]};
\addplot [xbar, fill=cblue, mark=none, draw=none, bar shift=0]
coordinates {
    (106.05, 1) [106 min 3 s]
    (60.27, 2)  [60 min 16 s]
    (42.5, 4)   [42 min 30 s]
    (75.85, 5)  [75 min 51 s]
    (55.2, 6)   [55 min 12 s]
    (55.77, 7)  [55 min 46 s]
    (76.45, 8)  [76 min 27 s]
};

\addplot [
    xbar, fill=cblue, mark=none, draw=none, bar shift=0,
    nodes near coords,
    nodes near coords style={
        font=\scriptsize,
        anchor=east,
        xshift=-2pt
    }
]
coordinates {
    (145.55, 3) [145 min 33 s]
};

            \addplot [cred, thick, dashed, line width=1pt] coordinates {(77.2, -1) (77.2, 9)};
            \node at (axis cs:78,0)[anchor=west,yshift=0pt,font=\scriptsize,text=cred]{\textbf{Participants Avg.: 77 min 12 s}};
            
        \end{axis}
    \end{tikzpicture}
    \caption{\label{fig:human-study-time-consumed}Time consumed (in minutes) by the LLM and each participant during the study.}

    \begin{tikzpicture}
        \begin{axis}[
            width=0.9\linewidth,
            ymajorgrids,
            grid style={dashed, line width=1pt},
            axis line style={draw=none},
            xtick style={draw=none},
            ytick style={draw=none},
            bar width=0.26cm,
            xmin=0.5,
            xmax=8.5,
            ymin=0,
            ymax=100,
            xlabel={Participant ID},
            ylabel={Agreement (\%)},
            ylabel style={yshift=-5px},
            xtick={1,2,3,4,5,6,7,8},
            ytick={0,20,40,60,80,100},
            legend style={
                at={(0.5,1)},
                anchor=south,
                font=\fontsize{7.5}{10}\selectfont,
                legend columns=2,
                draw=none,
                nodes={inner xsep=5pt},
                /tikz/every even column/.append style={column sep=8pt}
            },
            legend image code/.code={
                \draw[#1,draw=none] (0cm,-0.1cm) rectangle (0.2cm,0.1cm);
            }
        ]
            \addplot+[ybar, fill=cblue, mark=none, draw=none, bar shift=-0.15cm]  coordinates {(1, 80.0) (2, 80.0) (3, 88.0) (4, 88.0) (5, 84.0) (6, 84.0) (7, 92.0) (8, 64.0)};
            \addlegendentry{Participant $U_{\text{idx}}$}

            \addplot+[ybar, fill=cred, mark=none, draw=none, bar shift=0.15cm]  coordinates {(1, 88.0) (2, 84.0) (3, 84.0) (4, 84.0) (5, 88.0) (6, 88.0) (7, 88.0) (8, 84.0)};
            \addlegendentry{LLM-as-a-Judge}

            \addplot [black, thick, dashed, line width=1.2pt] coordinates {(0, 82.5) (9, 82.5)};
            \node at (axis cs:4.5,86)[anchor=south,yshift=0pt,font=\scriptsize,text=black]{\textbf{Participants Average: 82.5\%}};
        \end{axis}
    \end{tikzpicture}
    \caption{\label{fig:human-study-agreement}Results of the leave-one-annotator-out experiments comparing agreement between individual annotators and majority decisions. For each test case, we evaluate the agreement between a single annotator (either a human participant or LLM-as-a-Judge) and the majority vote of the remaining $N-1$ participants. The majority vote serves as the consolidated judgment from the excluded annotators.}

    \begin{tikzpicture}
        \begin{axis}[
            width=0.9\linewidth,
            xbar,
            xmajorgrids,
            grid style={dashed, line width=1pt},
            axis line style={draw=none},
            xtick style={draw=none},
            ytick style={draw=none},
            bar width=0.25cm,
            xmin=0,
            xmax=100,
            ylabel={Participant ID},
            xlabel={Agreement (\%)},
            ytick={1,2,3,4,5,6,7,8},
            xtick={0,20,40,60,80,100},
            legend style={
                at={(0.5,1.0)},
                anchor=south,
                font=\fontsize{7.5}{10}\selectfont,
                legend columns=2,
                draw=none,
                /tikz/every even column/.append style={column sep=5pt}
            },
            legend cell align=left,
            legend image code/.code={
                \draw[#1,draw=none] (0cm,-0.1cm) rectangle (0.2cm,0.1cm);
            }
        ]
            \addlegendimage{fill=cblue}
            \addlegendentry{Agree (163 / 200)}
            
            \addlegendimage{fill=light_blue}
            \addlegendentry{Accept → Reject (19 / 200)}
            
            \addlegendimage{fill=cred}
            \addlegendentry{Disapprove (8 / 200)}
            
            \addlegendimage{fill=light_yellow}
            \addlegendentry{Reject → Accept (10 / 200)}

            \addplot+[cblue, xbar stacked, bar shift=0] coordinates {(84.0, 1) (80.0, 2) (80.0, 3) (96.0, 4) (80.0, 5) (80.0, 6) (88.0, 7) (64.0, 8)};

            \addplot+[light_blue, xbar stacked, bar shift=0] coordinates {(16.0, 1) (0.0, 2) (4.0, 3) (0.0, 4) (12.0, 5) (16.0, 6) (8.0, 7) (20.0, 8)};

            \addplot+[light_yellow, xbar stacked, bar shift=0] coordinates {(0.0, 1) (8.0, 2) (0.0, 3) (4.0, 4) (8.0, 5) (4.0, 6) (4.0, 7) (12.0, 8)};

            \addplot+[cred, xbar stacked, draw=gray, line width=0.8pt, bar shift=0] coordinates {(0.0, 1) (12.0, 2) (16.0, 3) (0.0, 4) (0.0, 5) (0.0, 6) (0.0, 7) (4.0, 8)};
            
            \addplot[xbar, draw=black, line width=0.8pt, bar shift=0] coordinates {(100.0, 1) (88.0, 2) (84.0, 3) (100.0, 4) (100.0, 5) (100.0, 6) (100.0, 7) (96.0, 8)};
        \end{axis}
    \end{tikzpicture}
    \caption{\label{fig:human-study-proportion}Participants' agreement with the LLM before and after receiving the LLM's explanations. }

\end{figure}

\begin{figure}[t]
    \centering
    \captionsetup{skip=0pt}
    \includegraphics[width=\columnwidth]{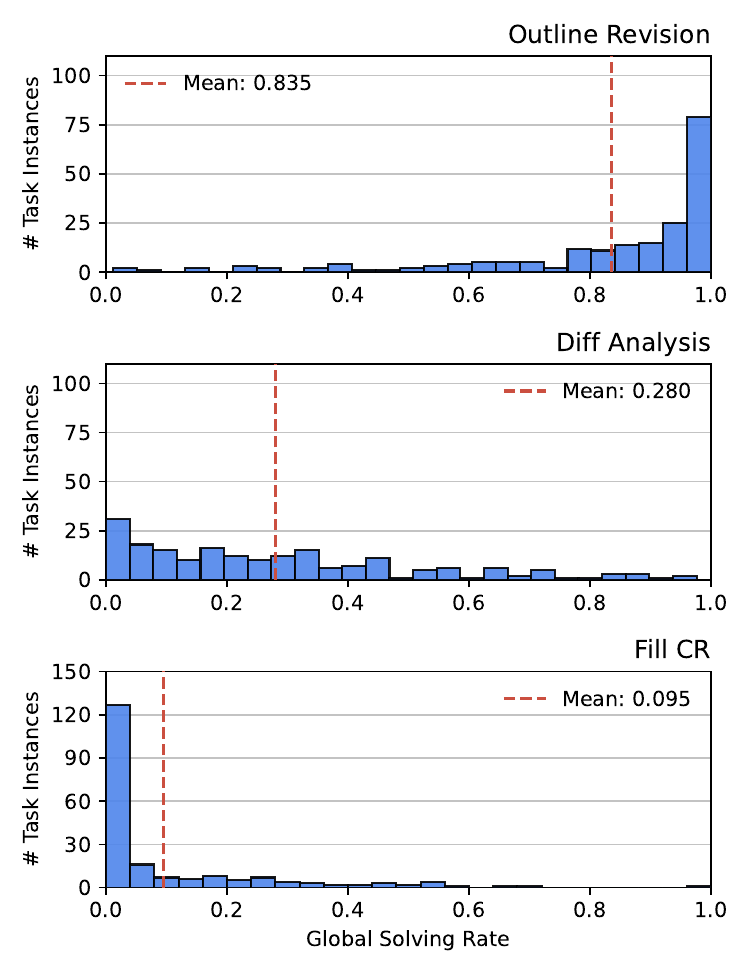}
    \caption{Solving rates of all models on the three tasks.}

    \label{fig:solving-rates-of-benchmark}
\end{figure}

The 25 LLM responses used in the human study belong to \texttt{GPT-4o} and a preview version of \ourmodel for \diffanalysistext and \fillcrtext. 
We randomly sample 12 acceptable and 13 unacceptable samples based on the LLM-as-a-Judge to ensure representativeness.
We collected a total of $8 * 25 * 2= 400$ responses, which are presented in~\Cref{tab:human-study-raw-data}. 
The time consumed by the LLM and each participant during the study is shown in~\Cref{fig:human-study-time-consumed}. 
The LLM completed the task in 36 seconds, significantly faster than the participants, who averaged 77 minutes and 12 seconds.
This suggests the unavailability of a large-scale study with human judgments and highlights the LLM-as-a-Judge's potential for time-efficient automation in evaluating LLM responses~\citep{zheng2024llm-judge-fastchat, chen2024alpagasus-data-cleaning-ref-1, ye2024justice-or-prejudice-LLM-as-a-Judge-reference-aware-ref}. 
The percentage of agreement between each participant and the LLM-as-a-Judge with the remaining participants is shown in~\Cref{fig:human-study-agreement}. 
For example, we compare Participant 1 and LLM-as-a-Judge with the consensus of Participants 2-8.
The LLM generally achieved a higher agreement rate, comparable to the participants' average of approximately 82.5\%. 
This indicates that the LLM's judgments align closely with participant consensus, supporting its reliability as an automated judge. 
\Cref{fig:human-study-proportion} presents participants' agreement with the LLM before and after receiving its explanations. 
Of 200 decisions, 163 consistently align with the LLM. 
Notably, 19 cases shifted from approval to rejection after the explanation, 10 shifted the opposite way, and 8 remained disapproved.
These patterns suggest that human participants and the LLM-as-a-Judge may hold different judgment criteria, which are effectively calibrated through the explanations provided in the judgment approval test.
These findings demonstrate our finally instantiated LLM-as-a-Judge's reliability in automating the evaluation of model answers at a notable expert level.

\begin{figure*}
    \centering
    \includegraphics[width=\linewidth]{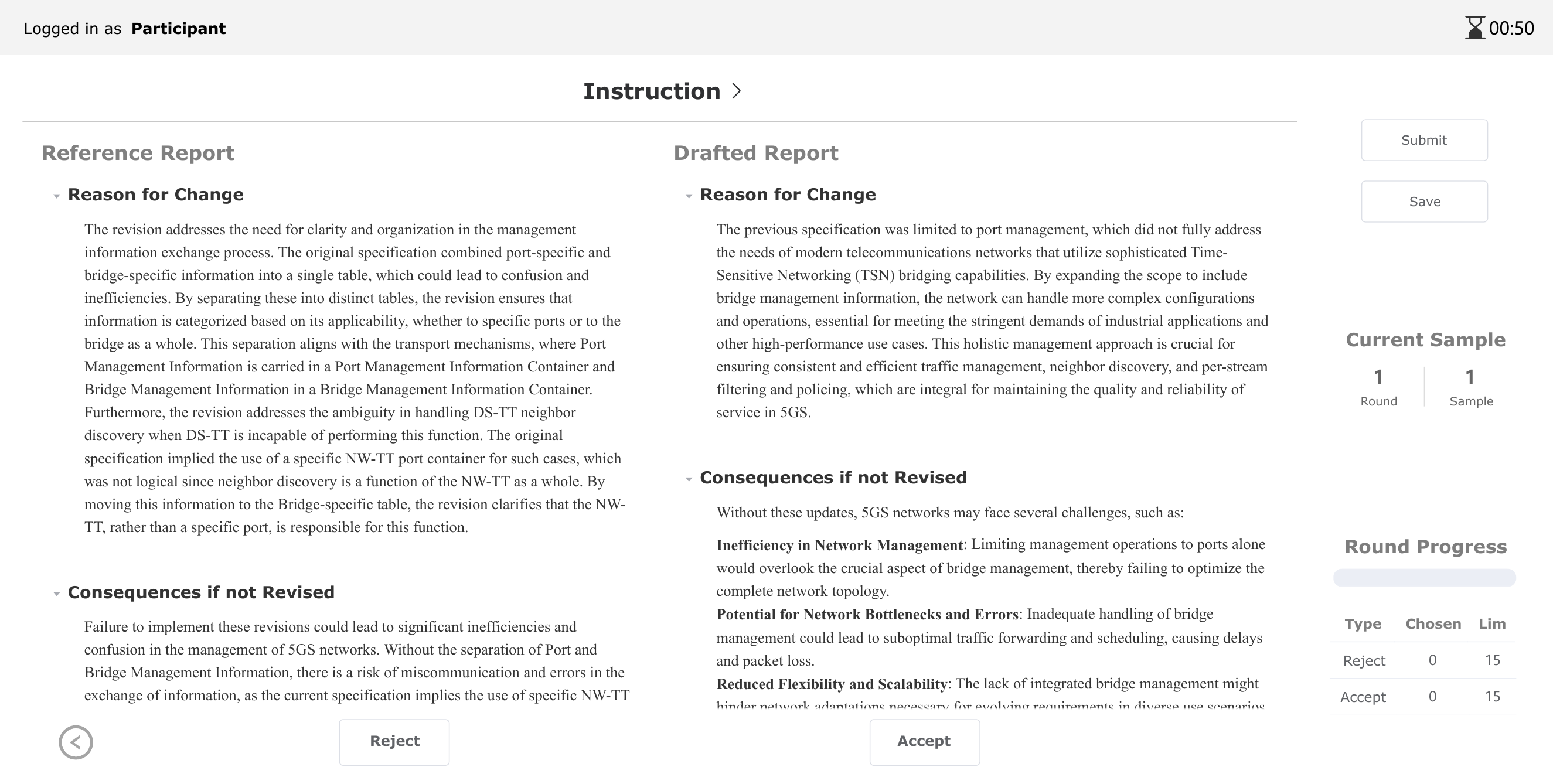}
    \caption{System snapshot of the first round of human study, Alignment Test.}
    \label{fig:human-study-system-snapshot-1}

    \centering
    \includegraphics[width=\linewidth]{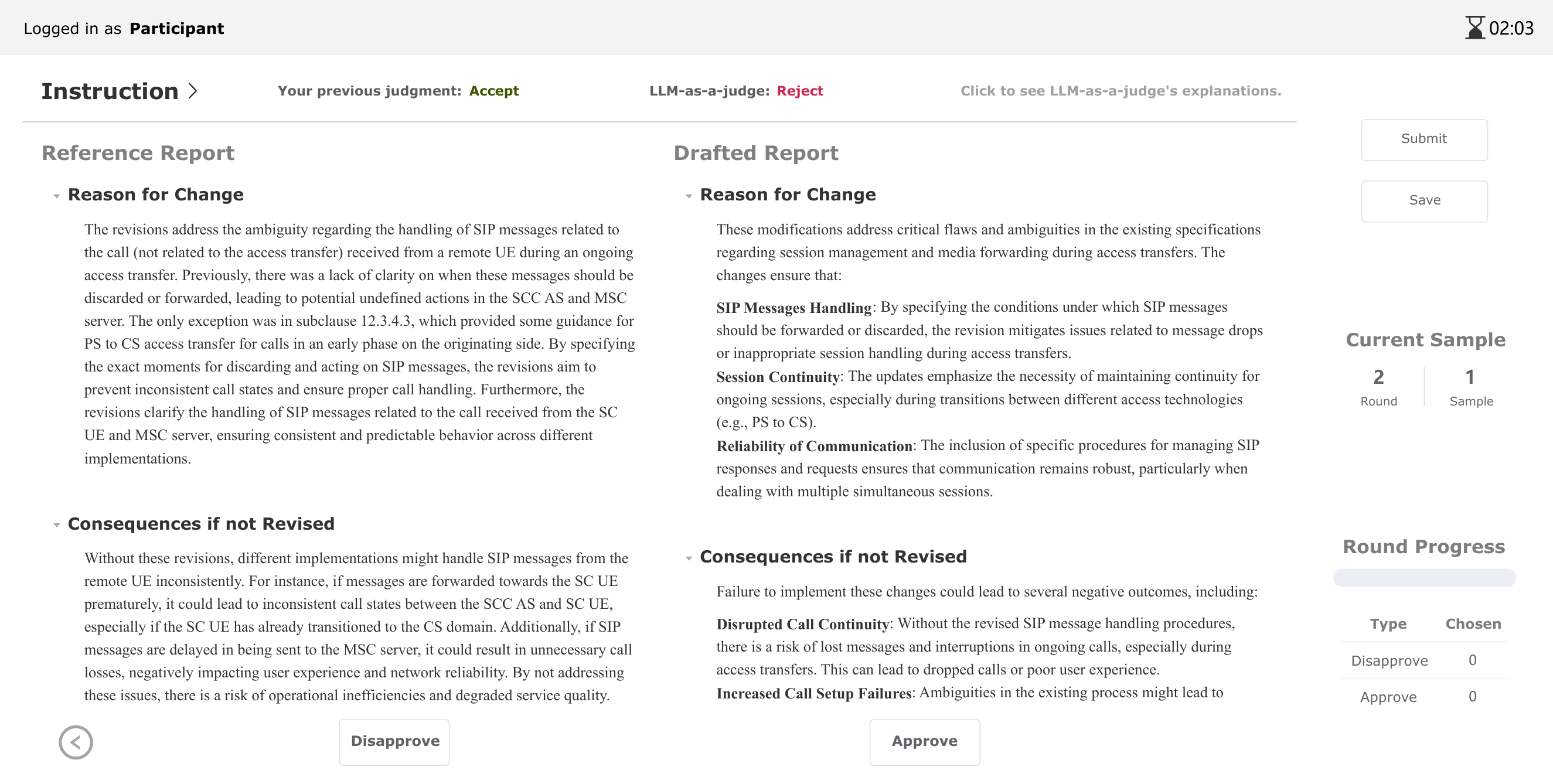}
    \caption{System snapshot of the second round of human study, Judgment Approval Test.}
    \label{fig:human-study-system-snapshot-2}
    
\end{figure*}

\clearpage
\onecolumn

\section{Prompts Used and Example Artifacts}\label{appx:examples-and-prompts}
\FloatBarrier

\begin{promptbox}[box:fill-cr-task-instruction]{\fillcrtitle}
{\small
You are a cellular network protocol expert. Given a segment of the 3GPP specifications, you should envision what bad things may happen when following the statements, and analyze its potential design weakness. Then, you prepare a change request, which should include:\\
1. REASON FOR CHANGE: Explain why the identified flaws need to be addressed.\\
2. SUMMARY OF CHANGE: Provide a summary of the necessary changes to the specifications.\\
3. CONSEQUENCES IF NOT REVISED: Describe potential negative impacts if the proposed changes are not made.\\
\\
You should avoid missing important statements and try your best to return detailed responses rich in reasoning.
}
\end{promptbox}

\begin{promptbox}[box:enrich-answer-rationale]{Rationale Augmentation}
{\small
You will be given a task instance composed of TASK INSTRUCTION, USER QUERY, and ASSISTANT RESPONSE. Your task is to revise the ASSISTANT RESPONSE by adding reasoning contents to it. The reasoning contents should explain how the response was generated and act as chain of thoughts for reaching the responses.

Note that\\
- The task will be related to network protocols, and you should leverage your knowledge in this domain.\\
- The revised response should be coherent with the original response.\\
- The revised response should perfectly fit the TASK INSTRUCTION and USER QUERY.\\
- The revised response should be informative and helpful to the user.\\
- The revised response should be rich in thoughts and smooth in logic.\\
- The revised response should be fruitful in educating other assistants.\\
- You should not alter the original response format.\\
- You should only return the revised response, which can directly replace the original response.\\

\# TASK INSTRUCTION\\
\textcolor{blue}{\{\}}\\
\# USER QUERY\\
\textcolor{blue}{\{\}}\\
\# ASSISTANT RESPONSE\\
\textcolor{blue}{\{\}}
}
\end{promptbox}

\begin{promptbox}[box:evaluate-security-relatedness]{Evaluating Security Relevance of CR}
{\small
You will be given a reasoning segment concerning analyzing problems of cellular network protocol. You should determine whether the implied problem is high-risk, meaning that it strongly relates to security, user privacy, attacks, or any threats to normal service. You should meticulously analyze the given task instance and end up with a judgment. If the problem discussed by the instance is high-risk, you should finally respond with 'High-Risk'; otherwise, respond with 'Low-Risk'.
\\ \\
$>>>$ REASON FOR CHANGE\\
\textcolor{blue}{\{\}}\\
$>>>$ CONSEQUENCES IF NOT REVISED\\
\textcolor{blue}{\{\}}
}
\end{promptbox}

\begin{promptbox}[box:locate-educational-samples]{Distinguishing Educational Task Instances}
{\small
You will be given a task instance concerning analyzing problems of cellular network protocol. The sample is composed of three parts, SYSTEM INSTRUCTION, USER QUERY, and ASSISTANT RESPONSE. You should determine its educational value for a student whose goal is to learn reasoning about the potential flaws in cellular network protocols. You should consider the following aspects: \\
\\ 
1. The ASSISTANT RESPONSE is self-contained, meaning that it does not depend on other external information too much and is easy to understand without additional context. \\
2. The USER QUERY is meaningful and is a suitable analytic target under the SYSTEM INSTRUCTION. \\
3. The ASSISTANT RESPONSE is logically coherent and conforms well to the requirements of the SYSTEM INSTRUCTION. \\
4. The ASSISTANT RESPONSE is informative and provides a detailed analysis of the USER QUERY. 
\\ \\
You should analyze the given task instance and end up with a judgement. DO NOT make your response too verbose. If it is educational, meaning that it is helpful for a student to learn reasoning about the potential flaws in cellular network protocols, you should finally respond with 'Educational'; otherwise, respond with 'Non-educational'.\\

\# TASK INSTRUCTION\\
\textcolor{blue}{\{\}}\\
\# USER QUERY\\
\textcolor{blue}{\{\}}\\
\# ASSISTANT RESPONSE\\
\textcolor{blue}{\{\}}
}
\end{promptbox}

\begin{promptbox}[box:evaluate-fill-cr]{Evaluating \fillcrtitle}
{\small
You are given two reports concerning a weakness analysis of the 3GPP protocol: the Reference Report (which should considered **acceptable** in quality of understanding the protocol problems) and the Drafted Report. Both reports aim to explain the weaknesses and reasons for certain revisions. Your task is to evaluate the claim the Drafted Report identifies the hidden problems in the protocol, compared to the Reference Report.\\

Conclude your evaluation with a judgment score (s) from:\\

* **-2 (Strongly Disagree):** The Drafted Report contains significant misunderstandings, misrepresents the issues, or omits critical weaknesses.\\
* **-1 (Weakly Disagree):** The Drafted Report shows a partial understanding but includes notable errors or omissions in identifying the weaknesses.\\
* **0 (Neutral):** The Drafted Report identifies the main weaknesses correctly but lacks depth or accuracy in some areas.\\
* **1 (Weakly Agree):** The Drafted Report largely understands the weaknesses but has minor discrepancies.\\
* **2 (Strongly Agree):** The Drafted Report demonstrates a near-perfect understanding of the potential problem, with only trivial deviations from the Reference Report.\\

Note that:\\

* The Reference Report provides a basic understanding of the protocol problems. The Reference Report is not perfect. So the Drafted Report does not need to match exactly with the Reference Report.\\
* You should focus on the protocol problems only. Ignore information unrelated to protocol problems in Reference Report, \eg reference to other documents.\\
* Superficial reports, those with speculative analysis, and those that lack focus should be rated lower. In contrast, reports that are decisive, informative, and facilitate further investigation by human experts are preferred.\\
* Focus on the content and the understanding of the protocol issues, not on the presentation or formatting.\\

Do not do anything else other than scoring. Only the final score (x) should be returned in the form of `s: x`.\\

\#\# Reference Report\\
\textcolor{blue}{\{\}}\\
\#\# Drafted Report\\
\textcolor{blue}{\{\}}
}
\end{promptbox}

\begin{promptbox}[box:gpt-verifier]{Weakness-to-Attack Verification}
{\small
You are an expert in verifying the correctness of a vulnerability analysis. You must be extremely rigorous and thorough in your verification process. Analyze and verify whether a given vulnerability analysis ($V$) enables an attacker, operating within a specified threat model ($T$), to execute the described attack ($A$). In other words, whether $V$ + $T$ -> $A$. Your verification must be comprehensive and consider:\\
1. Logical completeness - Are all necessary steps and conditions accounted for?\\
2. Technical accuracy - Are the technical details precise and correct?\\
3. Exploitability - Can the vulnerability be exploited within the constraints of the threat model?\\
4. Attack feasibility - Does the analysis conclusively demonstrate the attack's viability?\\
\\
Note that both $T$ and $A$ are trustworthy and you should evaluate the vulnerability analysis based on them.\\
\\
1. Vulnerability Analysis ($V$): Identified weaknesses from the specification\\
2. Threat Model ($T$): Attacker capabilities and assumptions\\
3. Attack Description ($A$): Attack procedure and implications\\
\\
ANALYSIS SECTIONS\\
----------------\\
\# Vulnerability Analysis\\
\textcolor{blue}{\{\}}\\
\\
\# Threat Model\\
\textcolor{blue}{\{\}}\\
\\
\# Attack Description\\
\textcolor{blue}{\{\}}\\
\\
OUTPUT FORMAT\\
------------\\
Please provide your detailed analysis in the following XML format:\\
<result>\\
    <correctness\_reason>\\
        Provide a thorough analysis addressing:\\
        1. Logical completeness of the vulnerability analysis\\
        2. Technical accuracy of all claims and assumptions\\
        3. Whether the vulnerability can be exploited given the threat model constraints\\
        4. Clear demonstration of attack feasibility\\
        5. Any gaps or inconsistencies found\\
        \\
        Support your conclusion with specific references to the input components.\\
    </correctness\_reason>\\
    <correctness>true/false</correctness>\\
</result>\\
\\
Note: Default to false if there is ANY uncertainty or gaps in the logical chain from vulnerability to attack.
}
\end{promptbox}

\begin{examplebox}[example:fill-cr-1]{\fillcrtitle Task in \oureval (Index: 29, CR: S3-190376, Design Flaw), Part: 1/2}
{\small
{\Large\textbf{Task Instruction}}\\
You are a cellular network protocol expert. Given a segment of the 3GPP specifications, you should envision what bad things may happen when following the statements, and analyze its potential design weakness. Then, you prepare a change request, which should include:\\
1. REASON FOR CHANGE: Explain why the identified flaws need to be addressed.\\
2. SUMMARY OF CHANGE: Provide a summary of the necessary changes to the specifications.\\
3. CONSEQUENCES IF NOT REVISED: Describe potential negative impacts if the proposed changes are not made.\\
\\
You should avoid missing important statements and try your best to return detailed responses rich in reasoning.\\

\par\vspace{-0.5em} %
\hrule height 1pt %
\par\vspace{0.5em} %
{\Large\textbf{Test Case}}\\
$>>>$ Original Specification Statements:\\

6.3.3 Authentication and key agreement\\
The purpose of this procedure is to authenticate the user and establish a new pair of cipher and integrity keys between the VLR/SGSN and the USIM. During the authentication, the USIM verifies the freshness of the authentication vector that is used.\\
\\
Figure 8: Successful UMTS Authentication and Key Agreement\\
The VLR/SGSN invokes the procedure by selecting the next unused authentication vector from the ordered array of authentication vectors in the VLR/SGSN database. Authentication vectors in a particular node are used on a first-in / first-out basis. The VLR/SGSN sends to the USIM the random challenge RAND and an authentication token for network authentication AUTN from the selected authentication vector.\\
Upon receipt the user proceeds as shown in Figure 9.\\
\\
Figure 9: User authentication function in the USIM\\
Upon receipt of RAND and AUTN the USIM first computes the anonymity key AK = f5K (RAND) and retrieves the sequence number SQN = (SQN  AK)  AK.\\
Next the USIM computes XMAC = f1K (SQN || RAND || AMF) and compares this with MAC which is included in AUTN. If they are different, the user sends an authentication failure message  back to the VLR/SGSN with an indication of the cause and the user abandons the procedure. In this case, VLR/SGSN shall initiate an Authentication Failure Report procedure towards the HLR as specified in section 6.3.6. VLR/SGSN may also decide to initiate a new identification and authentication procedure towards the user, cf. TS 24.008 [35].\\
Next the USIM verifies that the received sequence number SQN is in the correct range.\\
If the USIM considers the sequence number to be not in the correct range, it sends synchronisation failure back to the VLR/SGSN including an appropriate parameter, and abandons the procedure.\\
The synchronisation failure message contains the parameter AUTS. It is AUTS = Conc(SQNMS ) || MACS. Conc(SQNMS) = SQNMS   f5*K(RAND) is the concealed value of the counter SQNMS in the MS, and MACS = f1*K(SQNMS || RAND || AMF) where RAND is the random value received in the current user authentication request.  f1* is a message authentication code (MAC) function with the property that no valuable information can be inferred from the function values of f1* about those of f1, ... , f5, f5* and vice versa. f5* is the key generating function used to compute AK in re-synchronisation procedures with the property that no valuable information can be inferred from the function values of f5* about those of f1, f1*, f2, ... , f5 and vice versa.\\
\\
The AMF used to calculate MACS assumes a dummy value of all zeros so that it does not need to be transmitted in the clear in the re-synch message.
The construction of the parameter AUTS in shown in the following Figure 10:\\
\\
Figure 10: Construction of the parameter AUTS\\
If the sequence number is considered to be in the correct range however, the USIM computes RES = f2K (RAND) and includes this parameter in a user authentication response back to the VLR/SGSN. Finally the USIM computes the cipher key CK = f3K (RAND) and the integrity key IK = f4K (RAND). Note that if this is more efficient, RES, CK and IK could also be computed earlier at any time after receiving RAND. If the USIM also supports conversion function c3, it shall derive the 64-bit GSM cipher key Kc from the UMTS cipher/integrity keys CK and IK. UMTS keys are sent to the MS along with the derived 64-bit GSM key for UMTS-GSM interoperability purposes. USIM shall store original CK, IK until the next successful execution of AKA.\\
Upon receipt of user authentication response the VLR/SGSN compares RES with the expected response XRES from the selected authentication vector. If XRES equals RES then the authentication of the user has passed. The SGSN shall compute the 128-bit GSM ciphering key Kc128 according to annex B.5 if it is to use a 128-bit GSM ciphering algorithm. The VLR/MSC shall compute the 128-bit GSM ciphering key Kc128 according to annex B.5 if it signals a 128-bit GSM ciphering algorithm as a permitted GSM ciphering algorithm to the BSS. The VLR/SGSN also selects the appropriate cipher key CK and integrity key IK from the selected authentication vector. If XRES and RES are different, VLR/SGSN shall initiate an Authentication Failure Report procedure towards the HLR as specified in section 6.3.6. VLR/SGSN may also decide to initiate a new identification and authentication procedure towards the user, cf. TS 24.008 [35].\\

{\large\textcolor{gray}{To be continued in the next page \ding{216}}}
}
\end{examplebox}

\begin{examplebox}[example:fill-cr-2]{\fillcrtitle Task Instance in \oureval (Index: 29, CR: S3-190376, Design Flaw), Part: 2/2}
{\small
{\large\textcolor{gray}{\ding{216} Continued}}\\
Re-use and re-transmission of (RAND, AUTN)\\
The verification of the SQN by the USIM will cause the MS to reject an attempt by the VLR/SGSN to re-use a quintet to establish a particular UMTS security context more than once. In general therefore the VLR/SGSN shall use a quintet only once.\\
There is one exception however: in the event that the VLR/SGSN has sent out an authentication request using a particular quintet and does not receive a response message (authentication response or authentication failure) from the MS, it may re-transmit the authentication request using the same quintet. However, as soon as a response message arrives no further re-transmissions are allowed. If after the initial transmission or after a series of re-transmissions no response arrives, retransmissions may be abandoned. If retransmissions are abandoned then the VLR/SGSN shall delete the quintet. At the MS side, in order to allow this re-transmission without causing additional re-synchronisation procedures, the ME shall store for the PS domain (and optionally the CS domain) the last received RAND as well as the corresponding RES, CK and IK. If the USIM returned SRES and Kc (for GSM access), the ME shall store these values. When the ME receives an authentication request and discovers that a RAND is repeated, it shall re-transmit the response. The ME shall delete the stored values RAND, RES and SRES (if they exist) as soon as the 3G security mode command or the GSM cipher mode command is received by the ME or the connection is aborted. If the ME can handle the retransmission mechanism for CS domain then it shall be able to handle the retransmission for both PS and CS domain simultaneously.\\

6.3.5 Re-synchronisation procedure\\
A VLR/SGSN may send two types of authentication data requests to the HE/AuC, the (regular) one described in subsection 6.3.2 and one used in case of synchronisation failures, described in this subsection.\\
Upon receiving a synchronisation failure message from the user, the VLR/SGSN sends an authentication data request with a "synchronisation failure indication" to the HE/AuC, together with the parameters:\\
- RAND sent to the MS in the preceding user authentication request, and\\
- AUTS received by the VLR/SGSN in the response to that request, as described in subsection 6.3.3.\\
An VLR/SGSN will not react to unsolicited "synchronisation failure indication" messages from the MS.\\
The VLR/SGSN does not send new user authentication requests to the user before having received the response to its authentication data request from the HE/AuC (or before it is timed out).\\
When the HE/AuC receives an authentication data request with a "synchronisation failure indication" it acts as follows:\\
1. The HE/AuC retrieves SQNMS from Conc(SQNMS) by computing Conc(SQNMS)  f5*K(RAND).\\
2. The HE/AuC checks if SQNHE is in the correct range, i.e. if the next sequence number generated SQNHE using would be accepted by the USIM.\\
3. If SQNHE is in the correct range then the HE/AuC continues with step (6), otherwise it continues with step (4).\\
4. The HE/AuC verifies AUTS (cf. subsection 6.3.3).\\
5. If the verification is successful the HE/AuC resets the value of the counter SQNHE to SQNMS.\\
6. The HE/AuC sends an authentication data response with a new batch of authentication vectors to the VLR/SGSN. If the counter SQNHE was not reset then these authentication vectors can be taken from storage, otherwise they are newly generated after resetting SQNHE. In order to reduce the real-time computation burden on the HE/AuC, the HE/AuC may also send only a single authentication vector in the latter case.\\
Whenever the VLR/SGSN receives a new batch of authentication vectors from the HE/AuC in an authentication data response to an authentication data request with synchronisation failure indication it deletes the old ones for that user in the VLR/SGSN.\\
The user may now be authenticated based on a new authentication vector from the HE/AuC. Figure 12 shows how re-synchronisation is achieved by combining a user authentication request answered by a synchronisation failure message (as described in section 6.3.3) with an authentication data request with synchronisation failure indication answered by an authentication data response (as described in this section).\\
\\
Figure 12: Resynchronisation mechanism\\

\par\vspace{-0.5em} %
\hrule height 1pt %
\par\vspace{0.5em} %
{\Large\textbf{Reference Answer}}\\
$>>>$ REASON FOR CHANGE\\
The calculation of the keystream for concealing SQNMS in the case of re-sychronisation only has RAND as a non-key input and hence the same keystream may be used to protect different SQNMSs if an attacker replays an authentication challenge. This leaks at least the xor between the different SQNMS values and it has been shown how this leak can be used to derive the LSBs of SQNMS.\\
\\
$>>>$ SUMMARY OF CHANGE\\
Add MAC-S as an input parameter to claculate the keystream at re-synchronisation in order to get a differeent keystream for each SQNMS value.\\
\\
$>>>$ CONSEQUENCES IF NOT REVISED\\
Keystream is repeated which should be against security design properties.
}
\end{examplebox}

\begin{examplebox}[example:fill-cr-3]{\fillcrtitle  Task Instance in \oureval (Index: 145, CR: C1-193185, Under-specification)}
{\small
{\Large\textbf{Task Instruction}}\\
{\large\textcolor{gray}{For display brevity, the repetitive instruction is omitted here.}}\\

\par\vspace{-0.5em} %
\hrule height 1pt %
\par\vspace{0.5em} %
{\Large\textbf{Test Case}}\\
$>>>$ Original Specification Statements:\\
5.4.1.2.4.2 EAP message reliable transport procedure initiation by the network\\
In order to initiate the EAP message reliable transport procedure, the AMF shall create an AUTHENTICATION REQUEST message.\\
The AMF shall set the EAP message IE of the AUTHENTICATION REQUEST message to the EAP-request message to be sent to the UE. The AMF shall set the ngKSI IE of the AUTHENTICATION REQUEST message to the ngKSI value selected in subclause 5.4.1.2.2.2 or subclause 5.4.1.2.3.1. In this release of specification, the AMF shall set the ABBA IE of the AUTHENTICATION REQUEST message  with the length of ABBA IE to 2 and the ABBA contents to be 2 octets in length with value 0000H as described in subclause 9.11.3.10.\\
The AMF shall send the AUTHENTICATION REQUEST message to the UE, and the AMF shall start timer T3560 (see example in figure 5.4.1.2.4.2.1).\\
Figure 5.4.1.2.4.2.1: EAP message reliable transport procedure
Upon receipt of an AUTHENTICATION REQUEST message with the EAP message IE, the UE handles the EAP message received in the EAP message IE and the ABBA of the AUTHENTICATION REQUEST message.\\
\\
5.4.1.3.2 Authentication initiation by the network\\
The network may initiate a 5G AKA based primary authentication and key agreement procedure for a UE in 5GMM-CONNECTED mode at any time. For restrictions applicable after handover or inter-system change to N1 mode in 5GMM-CONNECTED mode, see subclause 5.4.1.2.3.\\
The network initiates the 5G AKA based primary authentication and key agreement procedure by sending an AUTHENTICATION REQUEST message to the UE and starting the timer T3560 (see example in figure 5.4.1.3.2.1). The AUTHENTICATION REQUEST message shall contain the parameters necessary to calculate the authentication response (see 3GPP TS 33.501 [24]). This message shall include the ngKSI that will be used by the UE and AMF to identify the KAMF and the partial native security context that is created if the authentication is successful. This message shall also include the ABBA parameter. In this release of specification, the network shall set the length of ABBA IE to 2 and the ABBA contents to be 2 octets in length with value 0000H as described in subclause 9.11.3.10.\\
If an ngKSI is contained in an initial NAS message during a 5GMM procedure, the network shall include a different ngKSI value in the AUTHENTICATION REQUEST message when it initiates a 5G AKA based primary authentication and key agreement procedure.\\
Figure 5.4.1.3.2.1: 5G AKA based primary authentication and key agreement procedure\\
\\
9.11.3.10 ABBA\\
The purpose of the ABBA information element is to enable the bidding down protection of security features.\\
The ABBA information element is coded as shown in figure 9.11.3.10.1 and table 9.11.3.10.1.\\
The ABBA is a type 4 information element with a minimum length of 4 octets.\\
8 7 6 5 4 3 2 1\\
ABBA IEI octet 1\\
Length of ABBA contents octet 2\\
ABBA contents octet 3 octet n\\
Figure 9.11.3.10.1: ABBA information element\\
Table 9.11.3.10.1: ABBA information element\\
ABBA contents (octet 3-n): indicate set of security features defined for 5GS as described in 3GPP TS 33.501 [24]. \\
\par\vspace{-0.5em} %
\hrule height 1pt %
\par\vspace{0.5em} %
{\Large\textbf{Reference Answer}}\\
$>>>$ REASON FOR CHANGE\\
CT1 sent an LS to SA3 (see C1-191686) asking about the UE behavior when an ABBA parameter with a non-zero value, or with a length that is more than 2 octets, is received in Release 15. This document aims to specify the UE behavior for this case following the LS response from SA3.\\
\\
$>>>$ SUMMARY OF CHANGE\\
Specify that in case the UE receives an ABBA parameter with - a length of more than 2 octets, - a non-zero value the UE shall use the ABBA as received from the network. Interoperability analysis1) UE compliant with the previous specification version with an AMF that is compliant with this CRIf the UE gets an ABBA parameter with a value that is different from 0000H, the UE will send a 5GMM STATUS message. However, the AMF already handles a 5GMM STATUS message.2) UE compliant with this CR with an AMF that is compliant with the previous specification version the UE gets an ABBA parameter with a value that is different from 0000H, the UE will use the ABBA parameter as it is received. If for some reason the KAMF at the UE and the network is not the same, the integrity check fails during the security mode procedure. However, handling integrity check failures already exists and is not introduced by this CR. If the KAMF at the UE and the network is the same, then no errors will occur. This CR is backwards compatible.\\
\\
$>>>$ CONSEQUENCES IF NOT REVISED\\
The UE uses an ABBA that is different from what the network has used leading to security failure.
}
\end{examplebox}

\clearpage

\renewcommand{\arraystretch}{1.2} %

\begin{table*}[htbp]
    \centering
    \caption{\label{tab:cr-example-part1}Meta-information and expert rationales for the example Change Request (C1-172658\protect\footnotemark)}
    \small

    \resizebox{\linewidth}{!}{%
        \begin{tabular}{|m{0.18\linewidth}|m{0.15\linewidth}|m{0.1\linewidth}|m{0.25\linewidth}|m{0.15\linewidth}|}
            \hline
            24.301 & 2871 & 2 & \textbf{Current version} & 14.3.0 \\
            \hline
        \end{tabular}
    }

    \resizebox{\linewidth}{!}{%
        \begin{tabular}{|m{0.215\linewidth}|m{0.55\linewidth}|m{0.15\linewidth}|m{0.1\linewidth}|}
            \hline
            \textbf{Title} & Correction of Handling of MO Detach without Integrity Protection & \textbf{Date} & 2017-05-19 \\
            \hline
            \textbf{Category} &  F (Correction) & \textbf{Release} & Rel-14 \\
            \hline
        \end{tabular}
    }

    \resizebox{\linewidth}{!}{%
        \begin{tabular}{|m{0.2\linewidth}|m{0.8\linewidth}|}
 \hline
            \textbf{Reason for change} & {In Rel-8, when CT1 specified the rules for the handling of NAS messages that are not intergrity protected or fail the integrity check by the receiver, CT1 decided that a mobile originated Detach Request without integrity protection was to be treated by the MME, because it is one of the messages which "in certain situations … are sent by the UE before security can be activated".\newline
\newline
An additional justification was at that time that it did not appear very likely that someone would take the efforts to listen in on the NAS signalling in a cell and operate a manipulated UE just for the purpose of detaching other subscribers. Moreover, in the worst case this kind of DoS attack, which would prevent the UE from receiving MT services, would be detected and alleviated when the UE performed the next normal or periodic TAU (or RAU) or when the UE requested some MO service.\newline
\newline
Since then, the situation has changed, because \eg for UEs used for MTC/ CIoT, it may take longer to detect and repair the issue, as\newline
- periodic update timer values up to ~14 days can be negotiated between UE and network, and\newline
- some of these devices send MO user data with a frequency of once every few weeks,\newline
but on the other hand the UEs (\eg certain metering devices) may be required to stay attached in order to be reachable for the application server.
Additionally, due to the higher density of devices per cell (and higher number per MME), it has become easier to perform the attack successfully even by picking the S-TMSI values at random.\newline
\newline
As there are not so many cases where a UE might rightfully send a Detach Request without integrity protection, we suggest to modify the requirements for the MME: the MME should authenticate the subscriber if possible. If the authentication is not performed, \eg because the detach is due to "switch-off", or for any other reason, the MME may ignore the Detach Request and remain in state EMM-REGISTERED. For this case the MME can attempt to apply additional criteria before marking the subscriber as deregistered, \eg the MME may wait whether the UE is still performing periodic updating or whether it is still responding to paging when an MT user data packet arrives. \newline
\newline
(We found the following cases where a UE might rightfully send a Detach Request without integrity protection:\newline
1) the UE is attached for emergency bearer services and there is no shared EPS security context available, \eg due to lack of roaming agreement;\newline
2) due to user interaction an attach procedure is cancelled before the secure exchange of NAS messages has been established;\newline
3) a NAS COUNT wrap around occurred so that the current EPS security context can no longer be used.\newline
In principle it should be possible for the MME to determine whether any of these cases can apply when a Detach Request message failing the integrity check is received.)} \\
            \hline
            \textbf{Summary of change} & {Rules for the handling of a DETACH REQUEST message failing the integrity check are modified for the case when a current EPS security context exists and the secure exchange of NAS messages has not yet been established:\newline
- If it is not a detach request due to switch off, and the MME can initiate an authentication procedure, the MME should authenticate the subscriber before processing the detach request any further.\newline
- If it is a detach request due to switch off, or the MME does not initiate an authentication procedure for any other reason, the MME may ignore the detach request and remain in state EMM-REGISTERED. (The network can attempt to use additional criteria before marking the UE as EMM-DEREGISTERED.)
} \\
            \hline
            \textbf{Consequences if not approved} & {Risk of a DoS attack. UEs using an extended periodic update timer can become unreachable for paging for a long time, if Detach Request without integrity protection is always accepted when secure exchange of NAS messages has not yet been established.} \\
            \hline
        \end{tabular}
    }

\end{table*}

\footnotetext{\url{https://www.3gpp.org/ftp/tsg_ct/WG1_mm-cc-sm_ex-CN1/TSGC1_104_Zhangjiajie/Docs/C1-172658.zip}}

\begin{table*}[htbp]
\centering
\small
\caption{\label{tab:cr-example-part2}Specification revisions for the example Change Request (C1-172658)}
\resizebox{\linewidth}{!}{
\begin{tabular}{|>{\centering\arraybackslash}p{0.02\linewidth}|p{0.95\linewidth}|}
\hline
 & {4.4.4.3 Integrity checking of NAS signalling messages in the MME} \\
 & Except the messages listed below, no NAS signalling messages shall be processed by the receiving EMM entity in the MME or forwarded to the ESM entity, unless the secure exchange of NAS messages has been established for the NAS signalling connection: \\
 & \quad - EMM messages: \\
 & \qquad - ATTACH REQUEST; \\
 & \qquad - IDENTITY RESPONSE (if requested identification parameter is IMSI); \\
 & \qquad - AUTHENTICATION RESPONSE; \\
 & \qquad - AUTHENTICATION FAILURE; \\
 & \qquad - SECURITY MODE REJECT; \\
 & \qquad - DETACH REQUEST; \\
& \qquad\qquad\qquad\qquad\qquad\qquad{\large\textcolor{gray}{The remaining unchanged clauses are omitted for brevity.}} \\
\text{[+]} & NOTE 2: The DETACH REQUEST message can be sent by the UE without integrity protection, \eg if the UE is attached for emergency bearer services and there is no shared EPS security context available, or if due to user interaction an attach procedure is cancelled before the secure exchange of NAS messages has been established. For these cases the network can attempt to use additional criteria (\eg whether the UE is subsequently still performing periodic tracking area updating or still responding to paging) before marking the UE as EMM-DEREGISTERED. \\
 & All ESM messages are integrity protected except a PDN CONNECTIVITY REQUEST message if it is sent piggybacked in ATTACH REQUEST message and NAS security is not activated. \\
 & Once a current EPS security context exists, until the secure exchange of NAS messages has been established for the NAS signalling connection, the receiving EMM entity in the MME shall process the following NAS signalling messages, even if the MAC included in the message fails the integrity check or cannot be verified, as the EPS security context is not available in the network: \\
 & \qquad - ATTACH REQUEST; \\
 & \qquad - IDENTITY RESPONSE (if requested identification parameter is IMSI); \\
 & \qquad - AUTHENTICATION RESPONSE; \\
 & \qquad - AUTHENTICATION FAILURE; \\
 & \qquad - SECURITY MODE REJECT; \\
\text{[-]} & \qquad - DETACH REQUEST (if sent before security has been activated); \\
\text{[+]} & \qquad - DETACH REQUEST; \\
 & \qquad - DETACH ACCEPT; \\
 & \qquad - TRACKING AREA UPDATE REQUEST; \\
 & \qquad - SERVICE REQUEST; \\
 & \qquad - EXTENDED SERVICE REQUEST; \\
 & \qquad - CONTROL PLANE SERVICE REQUEST. \\
 & NOTE 3: These messages are processed by the MME even when the MAC fails the integrity check or cannot be verified, as in certain situations they can be sent by the UE protected with an EPS security context that is no longer available in the network. \\
 & If an ATTACH REQUEST message fails the integrity check and it is not an attach request for emergency bearer services, the MME shall authenticate the subscriber before processing the attach request any further. For the case when the attach procedure is for emergency bearer services see subclause 5.5.1.2.3 and subclause 5.4.2.5. \\
\text{[+]} & If a DETACH REQUEST message fails the integrity check, the MME shall proceed as follows: \\
\text{[+]} & \quad - If it is not a detach request due to switch off, and the MME can initiate an authentication procedure, the MME should authenticate the subscriber before processing the detach request any further. \\
\text{[+]} & \quad - If it is a detach request due to switch off, or the MME does not initiate an authentication procedure for any other reason, the MME may ignore the detach request and remain in state EMM-REGISTERED. \\
 & NOTE 4: The network can attempt to use additional criteria (\eg whether the UE is subsequently still performing periodic tracking area updating or still responding to paging) before marking the UE as EMM-DEREGISTERED. \\
\text{[+]} & \qquad\qquad\qquad\qquad\qquad\qquad{\large\textcolor{gray}{The remaining unchanged clauses are omitted for brevity.}} \\
\hline
\end{tabular}
}
\end{table*}

\clearpage

\twocolumn

\begin{table*}[t]
    \centering

    \caption{Reported attacks in cellular networks.}
    \label{tab:reported-attacks}
    \resizebox{0.93\linewidth}{!}{
    \begin{tabular}{m{0.4\linewidth}>{\centering\arraybackslash}m{0.8\linewidth}}
        \toprule
        Attack Effects & Related Works\\
        \midrule
        IMSI/SUPI cracking & \citet{hussain2019privacy2019-side-channel} \\
        Traffic decryption & \citet{rupprecht2020call-me-maybe-traffic-decryption-and-eavesdropping} \textcolor{red}{} \\
        User tracking & \citet{kohls2019lost-traffic-encryption-traffic-fingerprinting-ref-1, hong2018guti-reallocation-demystified-user-tracking-ref-1, kotuliak2022ltrack-malicious-message-injection-ref-4} \\
        User presence identification & \citet{shaik2017practical-attacks-dos-ref-3-downgrading-ref-1, hussain2019privacy2019-side-channel, ludant20235g-sniffing-user-tracking-via-pattern-analysis, erni2022adaptover-malicious-message-injection-ref-2}  \\
        Device fingerprinting &  \citet{shaik2019exposing-device-capabilities-device-identification-ref-1, kotuliak2022ltrack-malicious-message-injection-ref-4, park2022doltest-specification-guided-negative-testing-judge-nlp-ref-1}\\
        Message/Service spoofing & \citet{kim2019touching-the-untouchables-dos-ref-2, lee2019your-president-spearking-attack, rupprecht2020imp4gt, park2022doltest-specification-guided-negative-testing-judge-nlp-ref-1} \citet{rupprecht2019breaking-lte-on-layer-2-dns-spoofing-and-website-spoofing-ref-1}\\
        Traffic fingerprinting & \citet{kohls2019lost-traffic-encryption-traffic-fingerprinting-ref-1, bae2022watching-the-watchers-video-identification-traffic-fingerprinting-ref-2} \\
        Denial of service & \citet{shaik2017practical-attacks-dos-ref-3-downgrading-ref-1, yu2019effects-of-control-signaling-dos-attack-ref-1, kim2019touching-the-untouchables-dos-ref-2, hussain2019privacy2019-side-channel, bitsikas2021handover-vulns, ludant2021sigunder-malicious-message-injection-ref-4,erni2022adaptover-malicious-message-injection-ref-2, akon2023formal-5g-access-control-nrf-hussain, chen2024taming-cellular-emergency-service, xing2024criticality-of-integrity-protection-of-5g-fronthaul-oran-dos-and-signaling-storm,bennett2024ransacked-fuzzing-core-ran-interface-lte-5g}  \\
        Downgrading to insecure versions &  \citet{shaik2017practical-attacks-dos-ref-3-downgrading-ref-1} \\
        Key re-installation & \citet{raza2021key-reinstallation-in-lte-ref-1} \\
        Malicious message injection & \citet{yang2019hiding-sigover-malicious-message-injection-ref-1, ludant2021sigunder-malicious-message-injection-ref-4, erni2022adaptover-malicious-message-injection-ref-2, kotuliak2022ltrack-malicious-message-injection-ref-4} \\
        Eavesdropping data communication & \citet{rupprecht2016testing-implementation-encryption-and-authentication,kim2019touching-the-untouchables-dos-ref-2, rupprecht2020call-me-maybe-traffic-decryption-and-eavesdropping, park2022doltest-specification-guided-negative-testing-judge-nlp-ref-1}  \\
        Exposing device capabilities & \citet{shaik2019exposing-device-capabilities-device-identification-ref-1}  \\
        Content Fingerprinting & \citet{kohls2019lost-traffic-encryption-traffic-fingerprinting-ref-1, bae2022watching-the-watchers-video-identification-traffic-fingerprinting-ref-2} \\
        Illegitimate access to services & \citet{rupprecht2020imp4gt, akon2023formal-5g-access-control-nrf-hussain} \\
        Unauthorized entry to secrets & \citet{akon2023formal-5g-access-control-nrf-hussain} \\
        Phishing legitimate users & \citet{kim2019touching-the-untouchables-dos-ref-2} \\
        BTS resource depletion & \citet{kim2019touching-the-untouchables-dos-ref-2} \\
        Free data service & \citet{li2015lyj-VoLTE,  chen2024taming-cellular-emergency-service} \\
        Signaling storm & \citet{xing2024criticality-of-integrity-protection-of-5g-fronthaul-oran-dos-and-signaling-storm} \\
        \bottomrule
    \end{tabular}}
\end{table*}

\section{Cellular Specification Weaknesses} \label{appx:spec-weaknesses}
Cellular specifications suffer from weaknesses, ranging from minor ambiguities and undefined behaviors to fundamental design flaws. 
While these weaknesses may go unnoticed under typical conditions, they can become threatening in specific scenarios.
A large number of CRs aim to address weaknesses in the specifications, motivating our research in this work.
Broadly speaking, these weaknesses can lead to various negative consequences, including performance degradation, interoperability failures, and security vulnerabilities.
In this work, we focus primarily on those weaknesses that pose security risks, among which the severest ones may be exploited by malicious entities to disrupt normal service operations.
While we focus on specification-level weaknesses, their implications are far-reaching. 
Design flaws within specifications lead to vulnerabilities in compliant implementations and thus propagate through the whole cellular network system.
Issues like under-specification lead to implementations and configurations that fail to meet essential requirements.
However, it would be unfair to place blame solely on specification drafters, particularly when observing the immense volume and complexity of cellular specifications. 
Current refinement practices depend on human experts to identify weaknesses and propose CRs, a labor-intensive approach that lacks a systematic evaluation framework.
These challenges highlight the critical need for automated tools capable of refining cellular specifications.

We provide a comprehensive survey of common specification weaknesses (\Cref{tab:common-spec-issues}) reported by previous academic works and associated attack vectors (\Cref{tab:reported-attacks}) in cellular networks.
Our survey reveals that numerous attacks against cellular networks have been proposed by exploiting unsafe designs and ambiguous drafts.
This demonstrates that specification weaknesses, if exploitable, can make significant impacts on cellular networks. 
However, it would be unfair to place blame solely on specification drafters, given the immense volume and complexity of the cellular network specification system.
Rather, it underscores the importance of systematic weakness analysis and motivates automatic tools that help refine cellular specifications.

\begin{table*}[t]
    \centering
    \caption{\label{tab:common-spec-issues}Common issues in cellular specifications.}
    \resizebox{0.9\textwidth}{!}{
        \begin{tabular}{p{4.5cm}C{11cm}}
        \toprule
        Specification weaknesses & Related works \\
        \midrule
            Design Flaws &  \citet{tu2014control-plane-lyj, shaik2017practical-attacks-dos-ref-3-downgrading-ref-1, yu2019effects-of-control-signaling-dos-attack-ref-1, shaik2019exposing-device-capabilities-device-identification-ref-1,  kim2019touching-the-untouchables-dos-ref-2, lee2019your-president-spearking-attack,hussain2019privacy2019-side-channel,  kotuliak2022ltrack-malicious-message-injection-ref-4,  ludant2021sigunder-malicious-message-injection-ref-4,chen2021bookworm,    bitsikas2021handover-vulns, bae2022watching-the-watchers-video-identification-traffic-fingerprinting-ref-2, borgaonkar2018new-privacy-threat-in-aka-user-tracking-ref-3} \\
             Underspecification & \citet{shaik2017practical-attacks-dos-ref-3-downgrading-ref-1, hong2018guti-reallocation-demystified-user-tracking-ref-1, basin20185g-aka-formal,  rupprecht2020call-me-maybe-traffic-decryption-and-eavesdropping, park2022doltest-specification-guided-negative-testing-judge-nlp-ref-1,akon2023formal-5g-access-control-nrf-hussain, xing2024criticality-of-integrity-protection-of-5g-fronthaul-oran-dos-and-signaling-storm} \\
             Undefined Behaviors & \citet{park2022doltest-specification-guided-negative-testing-judge-nlp-ref-1, klischies2023undefinedbehavior} \\
             Inconsistencies & \citet{park2022doltest-specification-guided-negative-testing-judge-nlp-ref-1, chen2024taming-cellular-emergency-service, rahman2024cellularlint} \\
         \bottomrule
        \end{tabular}
}
\end{table*}

\newcommand{\passed}{\ding{51}}
\newcommand{\failed}{\ding{55}}
\newcommand{\unfound}{-}
\newcommand{\implflaw}{\textcircled{I}}
\newcommand{\confflaw}{\textcircled{C}}
\newcommand{\notfoundattack}{\textcircled{N}}
\newcommand{\conditionalpassed}{\textcircled{$\checkmark$}}

\definecolor{customshadow}{RGB}{217, 232, 245} %

\newcommand{\shadowed}[1]{%
  \tcbox[size=fbox, on line, colback=rowblue, colframe=black!50, 
         boxrule=0.3mm, arc=1mm, shadow=0.2mm]{#1}%
}

\begin{table*}[h]
\centering
\caption{Evaluation of known attacks using \ourmodel (\shadowed{C}). Hermes results (\shadowed{H}) are self-reported in~\citet{al2024hermes}. We indicate the version where the flawed specification was identified. We use symbols \implflaw~(implementation flaw), \confflaw~(configuration flaw), and \notfoundattack~(non-deterministic).}
\label{tab:known-vulns}
\small
\resizebox{0.9\textwidth}{!}{
    \begin{tabular}{l p{0.5\textwidth}ccc}
    \toprule
    ID & Attack & Protocol & \shadowed{H} & \shadowed{C} \\
     \midrule
        1 & AUTH REJECT Attack~\citep{yu2019effects-of-control-signaling-dos-attack-ref-1} & 4G NAS (15.0.0) &  \passed & \passed \\
        2 & Blind DoS Attack~\citep{kim2019touching-the-untouchables-dos-ref-2} & 4G RRC (14.2.2) & \failed  & \passed \\
        3 & Cutting off the Device~\citep{hussain20195greasoner} & 5G NAS (16.2.0) & \failed  & \passed \\
        4 & Deletion of allowed CAG list~\citep{al2024hermes} & 5G NAS (17.8.0) & \passed  & \passed  \\
        5 & DoS with RRCSetupRequest attack~\citep{hussain20195greasoner}  & 5G RRC (15.5.1) & \failed & \passed \\
        6 & Denying all network services~\citep{shaik2017practical-attacks-dos-ref-3-downgrading-ref-1}  & 4G NAS (12.8.0) & \passed & \passed \\
        7 & Denying selected service~\citep{shaik2017practical-attacks-dos-ref-3-downgrading-ref-1}  & 4G NAS (12.8.0) & \failed & \passed \\
        8 & DETACH REQUEST attack~\citep{hussain2018lteinspector} & 4G NAS (12.8.0) &  \passed & \passed \\
        9 & Downgrade to non-LTE services~\citep{shaik2017practical-attacks-dos-ref-3-downgrading-ref-1} & 4G NAS (12.8.0) & \passed & \passed \\
        10 & Downgrade via ATTACH REJECT~\citep{shaik2017practical-attacks-dos-ref-3-downgrading-ref-1} & 4G NAS (12.8.0) & \passed & \passed \\
        11 & Energy Depletion with RRCSETUP~\citep{al2024hermes} & 5G RRC (17.0.0) & \passed  & \passed \\
        12 & Exposing NAS Sequence Number~\citep{hussain20195greasoner}  & 5G NAS (16.0.2) & \passed & \passed \\
        13 & Exposure of SQN~\citep{borgaonkar2018new-privacy-threat-in-aka-user-tracking-ref-3} & 3G AKA (15.0.0) & \passed & \passed \\
        14 & IMSI Catching~\citep{van2015defeating-imsi-catchers} & 4G NAS (12.7.0) & \passed & \passed \\
        15 & IMSI Cracking~\citep{hussain2019privacy2019-side-channel} & 4G RRC (15.0.0) & \failed & \passed \\
        16 & IMSI Cracking~\citep{hussain2019privacy2019-side-channel} & 5G NAS (15.0.0) & \failed & \passed \\
        17 & Incarceration with RRCRELEASE~\citep{hussain20195greasoner} & 5G RRC (15.5.1) & \passed & \passed \\
        18 & Installing Null Cipher/Integrity~\citep{hussain20195greasoner}  & 5G RRC (15.5.1) & \passed & \passed \\
        19 & Lullaby Attack~\citep{hussain20195greasoner} & 5G RRC (15.5.1) & \passed & \passed \\
        20 & Measurement report~\citep{shaik2017practical-attacks-dos-ref-3-downgrading-ref-1} & 4G RRC (12.3.0) & \failed & \passed  \\
        21 & NAS COUNT update attack~\citep{al2024hermes} & 5G NAS (16.4.0) & \passed & \passed \\
        22 & NAS Counter Reset~\citep{hussain20195greasoner}  & 5G NAS (16.0.2) & \passed &  \passed \\
        23 & Neutralizing TMSI Refreshment~\citep{hussain20195greasoner} & 5G NAS (16.2.0) & \failed & \passed \\
        24 & Paging channel hijacking~\citep{hussain2018lteinspector} & 4G RRC (12.5.0) & \failed & \passed \\
        25 & SERVICE REJECT attack~\citep{shaik2017practical-attacks-dos-ref-3-downgrading-ref-1} & 4G NAS (12.8.0) & \passed & \passed \\
        26 & Signaling DoS Attack~\citep{bassil2013effects-of-signaling-attacks-on-lte-networks} & 4G NAS (16.8.0) & \passed & \passed \\
        27 & SUCI Catching Vulnerability~\citep{chlosta20215g-suci-catcher} & 5G NAS (15.0.0) & \failed &  \passed \\
        28 & Synchronization Failure Attack~\citep{yu2019effects-of-control-signaling-dos-attack-ref-1} & 4G NAS (15.0.0) & \failed & \passed  \\
        29 & Uplink NAS Counter Desync~\citep{hussain20195greasoner} & 5G NAS (16.0.2) & \passed & \passed \\
        30 & 5G AKA DoS Attack~\citep{cao2019survey-security-aspects-3gpp-5g-networks} & 5G NAS (15.2.0) & \passed & \passed \\
        \midrule
        31 & AKA Bypass~\citep{kim2019touching-the-untouchables-dos-ref-2} & 5G RRC (\implflaw) & \failed & \unfound \\
        32 & EMM Information Vulnerability~\citep{park2016white-rabbit-in-mobile-effect-of-unsecured-clock-source-in-smartphones} & 4G NAS (\implflaw) & \passed & \unfound  \\
        33 & Impersonation attack~\citep{chlosta2019lte-configuration-failure} & 4G NAS (\confflaw) & \failed & \unfound \\
        34 & Malformed Identity Request~\citep{michau2016how-to-not-break-lte-crypto} & 4G NAS (\implflaw) & \failed & \unfound \\
        35 & RLF report~\citep{shaik2017practical-attacks-dos-ref-3-downgrading-ref-1} & 5G RRC (\implflaw) & \passed & \unfound \\
        36 & S-TMSI Catching~\citep{kim2019touching-the-untouchables-dos-ref-2} & 4G NAS (\notfoundattack) & \passed & \unfound \\
         \bottomrule
    \end{tabular}}

\end{table*}

\end{document}